\documentclass[preprint]{iucr}              

     \journalcode{J}

 \usepackage{mathtools}
  \usepackage{xcolor}
 \usepackage{algorithm}

\DeclarePairedDelimiter{\ceil}{\lceil}{\rceil}

\RequirePackage{amsmath}
\RequirePackage{xspace}
\RequirePackage{bbm}

\def\XS{\xspace}
\DeclareMathAlphabet{\mathb}{OML}{cmm}{b}{it}
\def\sbm#1{\ensuremath{\mathb{#1}}}                
\def\sbmm#1{\ensuremath{\boldsymbol{#1}}}

\def\scu#1{\ensuremath{\mathcal{#1\XS}}}           
 
\def\sbl#1{\ensuremath{\mathbbm{#1}}}

\def\Bb{{\sbm{B}}\XS}  \def\bb{{\sbm{b}}\XS}

  \def\eb{{\sbm{e}}\XS}

  \def\kb{{\sbm{k}}\XS}
  
  \def\mb{{\sbm{m}}\XS}
  \def\nb{{\sbm{n}}\XS}

  \def\qb{{\sbm{q}}\XS}
  \def\rb{{\sbm{r}}\XS}

\def\Fc{{\scu{F}}\XS}

\def\Cbb{{\sbl{C}}\XS}

\def\Rbb{{\sbl{R}}\XS}

\def\Zbb{{\sbl{Z}}\XS}

      \def\Deltab    {{\sbmm{\Delta}}\XS}

     \def\Lambdab   {{\sbmm{\Lambda}}\XS}
\def\mub         {{\sbmm{\mu}}\XS}

\def\psib        {{\sbmm{\psi}}\XS}        \def\Psib      {{\sbmm{\Psi}}\XS}

\def\eC{\Cbb}

\def\eR{\Rbb}
\def\eZ{\Zbb}

\begin{document}

\title{General approaches for shear-correcting coordinate
  transformations in  Bragg coherent diffraction imaging: Part II}
\shorttitle{Shear-correcting coordinate  transformations in BCDI: Part II}
\date{}

\author[a]{P.}{Li}
\author[b]{S.}{Maddali}
\author[c,d]{A.}{Pateras}
\author[b]{I.}{Calvo-Almazan}
\author[b]{S.O.}{Hruszkewycz}
\author[a]{V.}{Chamard}
\cauthor[a]{M.}{Allain}{marc.allain@fresnel.fr}{}

\aff[a]{Aix-Marseille Univ, CNRS, Centrale Marseille, Institute Fresnel, 13013 Marseille, \country{France}}
\aff[b]{Materials Science Division, Argonne National Laboratory, Lemont, IL 60439, \country{USA}}
\aff[c]{Center for Integrated Nanotechnologies, Los Alamos National Laboratory, Los Alamos, NM 87545, \country{USA}}
\aff[d]{Materials Science and Technology Division, Los Alamos National Laboratory, Los Alamos, NM 87545, USA}
\shortauthor{P. Li \textit{et al.}}

\begin{abstract}
X-ray Bragg coherent diffraction imaging has been demonstrated as a powerful three-dimensional (3D) microscopy approach for the investigation of sub-micrometer-scale crystalline particles. The approach is based on the measurement of a series of coherent Bragg diffraction intensity patterns that are numerically inverted to retrieve an image of the spatial distribution of relative phase and amplitude of the Bragg  structure factor of  the diffracting sample. This 3D information, which is collected through an angular rotation of the sample, 
is necessarily obtained in a non-orthogonal frame in Fourier space that must be eventually reconciled. 
To deal with this, the approach currently favored by practitioners (detailed in Part I) is to perform the entire inversion in conjugate non-orthogonal real and Fourier space frames, and to transform the 3D sample image into an orthogonal frame as a post-processing step for result analysis. 
In this article, which is a direct follow-up of Part I, we demonstrate two different transformation strategies that enable the entire inversion procedure of the measured data set to be performed in an orthogonal frame. 
	The new approaches described here build mathematical and numerical frameworks that apply to the cases of evenly and non-evenly sampled data along the direction of sample rotation (\textit{i.e.}, the rocking curve). 
The value of these methods is that they rely on and incorporate significantly more information about the experimental geometry into the design of the phase retrieval Fourier transformation than the strategy presented in Part I. 
Two important outcomes are 
1) that the resulting sample image is correctly interpreted in a shear-free frame, and 
2) physically realistic constraints of BCDi phase retrieval that are difficult to implement with current methods are easily incorporated. 
Computing scripts are also given to aid readers in the implementation of the proposed formalisms. 
\end{abstract}
\keyword{Bragg coherent diffraction imaging, Fourier synthesis, non-orthogonal Fourier sampling, coordinate
  transformation, shear correction, Bragg ptychography.}
\maketitle

\section{Introduction}

Coherent diffraction imaging (CDI) approaches based on x-ray Bragg
diffraction have emerged as valuable tools for materials
science, owing to their sensitivity to atomic
displacement fields, their three-dimensional (3D) imaging capability, their
high-spatial resolution \cite{robinson03}, and their suitability for
 non-destructive investigation
of complex material systems in various environments \cite{Ulvestad15}.  
These methods, including single-particle Bragg coherent diffraction imaging (BCDI) 
 \cite{williams03,pfeifer06}  and Bragg
ptychography \cite{godard11,Hruszkewycz12,Mastropietro17}, 
rely on the measurement of far-field x-ray coherent intensity patterns in the vicinity of a Bragg peak that result from 
a crystalline sample being illuminated with a coherent x-ray beam.
One unique aspect of BCDI as compared to forward-scattering 3D CDI approaches \cite{chapman10,Miao15}, is the way
the information is obtained. 
In a Bragg diffraction geometry, the 3D
information is gathered through a series of 2D measurements, which
 correspond to closely-spaced sequential parallel planar slices of 
the Fourier space 3D intensity pattern. Thus, the stacking of the 2D
measurements produces a 3D intensity data set that contains information about the 3D structural features of the diffracting sample. 
This data set is numerically inverted to yield a 3D real-space array that describes the sample structure.

However, one important consideration in BCDI is the fact that the directions of Fourier space sampling of a BCDI scan 
are necessarily non-orthogonal (as discussed in Part I).
This holds true for the cases when the data are obtained using an angular sample scan (along the rocking curve or RC) \cite{williams03}
or an incident beam energy scan \cite{cha16}.  
The inherent non-orthogonal nature of the Fourier space measurement has to be accounted for when interpreting the retrieved image of the 3D sample.
One strategy to deal with the non-orthogonal measurement frame is presented in Part I 
of this work, and is summarized here. 
Coherent far-field diffraction from a scatterer in the Bragg condition is the squared modulus of the Fourier transform of the complex-valued 3D scatterer~\cite{Takagi69,Vartanyants2001}, centered at the Bragg peak of interest.
The fact that measurement of such a Bragg peak is necessarily performed on a non-orthogonal basis in Fourier space
implies that the resultant real-space sample 
is likewise described in a conjugate sheared frame. 
Thus, a real space transformation, which we formally derive in Part I, must be applied after the completion of phase retrieval in order to visualize the  sample  in an orthogonal frame \cite{berenguer13,Yang2019}.
This approach is the one typically implemented in most of the Bragg
CDI literature published to date \cite{pfeifer06,Ulvestad15}.

The motivation for Part II is born of a realization that the approach of rectifying the frame of the sample after phase retrieval, though commonly implemented in BCDI, significantly limits the ability to incorporate constraints stemming from the physics of the experiment or from the geometry of the sample into the phase retrieval process. 
For example, the behavior of the commonly-utilized shrinkwrap algorithm \cite{Marchesini03} could be much more carefully controlled if the blurring kernel were not cast in dimension- and direction-agnostic ``pixel'' coordinates, as it is today, but rather in real space units that impact all facets and corners of the reconstructed object symmetrically in 3D.  
Similarly, efforts to date to account for partial coherence effects in BCDI data have treated the problem as an ad-hoc blind ``deblurring'' problem~\cite{Clark2012} rather than building in estimates  based on known de-cohering effects that can  be calculated for x-ray synchrotron beamlines. 
Additionally, when considering Bragg CDI methods more broadly, working in a sheared sample frame greatly complicates the description and placement of a 
localized beam in rocking-curve 3D Bragg ptychography methods~\cite{hill18,Hruszkewycz12}. 
Solutions to these, and other related problems, all hinge on a phase-retrieval description of the sample on an orthogonal real space reference frame onto which other experimental constraints and models map naturally.
A recent example, wherein such a strategy was utilized to determine the angular uncertainty of each measurement step in a rocking curve during the course of image reconstruction with phase retrieval~\cite{calvo-almazan19}, shows the potential of such a construction. 
We anticipate that further advances of this nature could be possible, provided that formalisms for experimental-geometry-aware Fourier transformations are developed and provided to the community. 

In this article we present the  framework  for two  computationally efficient Fourier
transformations that simultaneously offer a natural frame both for the sample (orthogonal real space frame) and for the 
data (non-orthogonal rocking curve sampling of Fourier space) that can be integrated into common phase retrieval algorithms. 
The first method is derived for the case where an evenly
sampled data set is obtained by typical rocking curve measurement methods. 
It is based on a physically-informed description of the conjugate relations
between the 3D real and Fourier spaces that is compatible with computationally efficient discrete Fourier transformations. 
The second  approach, which has been introduced and used in earlier Bragg CDI-related works
\cite{hruszkewycz17,hill18,calvo-almazan19}, 
is more flexible and is built on a concept that exploits the properties of Fourier slice projection. 
For the sake of clarity, in building our calculations, we adopt the relatively simple, though highly utilized case of symmetric x-ray diffraction with a two-circle diffractometer (sometimes referred to a symmetric $\theta$-$2\theta$ geometry).
The extension to the non-symmetric case involving more 
than two diffractometer angles, as was dealt with in Part I, is relatively
straightforward and is described in the Appendix.

\section{The coordinate transform and conjugation relation}
\label{secQSCT}
 
We briefly summarize the conjugate relation between coordinate representations of real and Fourier spaces.
We refer the reader to Part I for further details
\footnote{
We adopt the notations and conventions used 
in Part I, with a single adaptation: 
all frames introduced in part II are built with 
unit-norm vectors by default, therefore we drop the caret 
symbol '$\,\widehat{\cdot}\,$' in the unit-norm vector notation, 
\textit{e.g.,} the unit-norm vectors $\hat{\eb}$, $\hat{\tilde{\eb}}$, 
 $\hat{\kb}$ and $\hat{\tilde{\kb}}$ introduced in this section are 
unambiguously written hereafter $\eb$, $\tilde{\eb}$, $\kb$ and $\tilde{\kb}$, 
unless explicitly stated otherwise.
}.

We denote $\tilde{\rb} := [\tilde{r}_1~~\tilde{r}_2~~\tilde{r}_3]^T$ as a point
in real or direct space 
that is defined with respect to a frame $(\tilde{\eb}_1, \tilde{\eb}_2,
\tilde{\eb}_3)$. 
Here `$T$' denotes the matrix transpose.
Similarly, we denote  $\tilde{\qb} := [\tilde{q}_1~~\tilde{q}_2~~\tilde{q}_3]^T$ as a point
in Fourier space (\textit{i.e.,} associated with the
measurement)  defined with respect to a frame $(\tilde{\kb}_1, 
\tilde{\kb}_2, \tilde{\kb}_3)$. In addition, we define the Fourier pair
$\tilde{\psi} \rightleftharpoons   \tilde{\Psi}$ by
\begin{equation}
  \label{FT}
     \tilde{\Psi}(\tilde{\qb})  =  \int_{\eR^3 } \tilde{\psi}(\tilde{\rb})\,
     e^{-\iota 2\pi \tilde{\rb}^T \tilde{\qb}} \,\text{d} \tilde{\rb} 
\end{equation}
where $\iota = \sqrt{-1}$.
This relationship implies that $(\tilde{\eb}_1, \tilde{\eb}_2, \tilde{\eb}_3)$ and 
$(\tilde{\kb}_1, \tilde{\kb}_2, \tilde{\kb}_3)$ are dual, or conjugate, frames.
The field $\tilde{\Psi} $ can be represented in any other (Fourier-space) 
frame  $(\kb_1, \kb_2, \kb_3)$ through an
appropriate linear coordinate transformation. 
We can obtain such an alternative frame of representation on a new 
Fourier-space basis by applying a linear transformation:
\begin{equation}
  \label{basis_QS}
     \qb  =  \Bb_q \tilde{\qb}. 
\end{equation}
In this expression, $\Bb_q\in \eR^{3\times 3}$ is the ``original'' Fourier frame $(\tilde{\kb}_1, 
\tilde{\kb}_2, \tilde{\kb}_3)$ expressed in the ``new''  Fourier frame 
$(\kb_1, \kb_2, \kb_2)$. 
With a change of frame in $\qb$, also comes a change in the dual (real-space) 
basis frame $(\eb_1, \eb_2, \eb_3)$ in which the vector $\rb:=[r_1~~r_2~~r_3]^T$ is expressed. 
As in Fourier space, the re-framed real space variable $\rb$ is related to $\tilde{\rb}$ by a linear coordinate transform
\begin{equation}
  \label{basis_RS}
      \rb  =  \Bb_{r} \tilde{\rb},
\end{equation}
with $\Bb_{r}\in \eR^{3\times 3}$. For the sake of simplicity, from now on we 
adopt the simplified notation $\Bb_r \equiv \Bb$ in this paper. From Part I (see Sec.~2),  the real-space linear 
transformation and its Fourier-space counterpart are related by 
\begin{equation}
  \label{basis_RS_RQ}
       \Bb_{q} = \Bb^{-T}
\end{equation}
with $\Bb^{-T} = (\Bb^{T})^{-1} = (\Bb^{-1})^{T} $. 
Coordinate transformations in Fourier space can be implemented by the following relationships:
\begin{equation}
  \label{FT_2}
   \Psi(\qb) =     \text{det} (\Bb) \,\tilde{\Psi}(\Bb^T\qb)   
  \quad \Leftrightarrow\quad
    \tilde{\Psi}(\tilde{\qb}) =    \frac{1}{\text{det} (\Bb)}  \Psi(\Bb^{-T} \tilde{\qb}). 
\end{equation}
As shown in Part I, because of the Fourier relationship between
 ${\psi} \rightleftharpoons   {\Psi}$, the means of coordinate transformation in real space are determined to be:
 $\psi(\rb):=\tilde{\psi}(\Bb^{-1}\rb)$. 
 The functions $\tilde{\psi}$ 
and ${\psi}$ actually represent the 
\textit{same} real-space object but in 
different frames.
The same holds true for $\tilde{\Psi}$ and $\Psi$ regarding the Fourier space representation. 
The equations in \eqref{FT_2} are therefore useful 
as they describe how any measurement in $(\tilde{\kb}_1, \tilde{\kb}_2,
\tilde{\kb}_3)$ can be mapped to an alternative frame $({\kb}_1,
{\kb}_2, {\kb}_3)$, and \textit{vice versa}. 
In particular, in the subsequent sections, we show the means by which to construct BCDI-compatible transformations that work within an alternative, orthogonal frame that is built upon the specific geometric considerations of the experiment.

\section{Application to BCDI: from a non-orthogonal to an orthogonal frame}
\label{mapping}
\begin{figure}
\centering
\includegraphics[width=.95\textwidth]{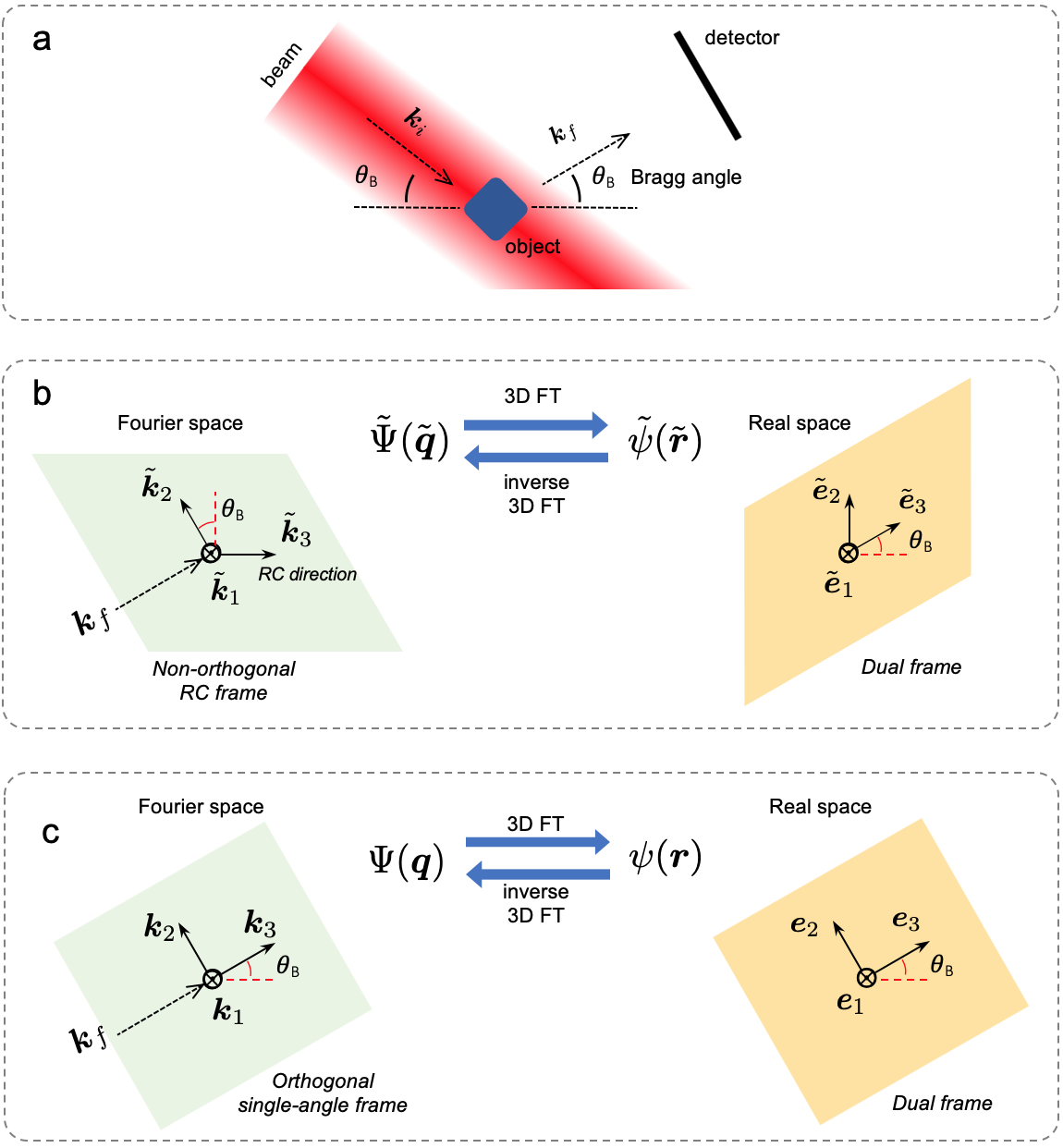}
\caption{Bragg coherent diffraction: geometry and frame definitions. 
(a) During a typical x-ray coherent diffraction imaging experiment 
the incident vector $\kb_i$  and the exit vector $\kb_f$ are in Bragg 
condition: the vector pair ($\kb_i, \kb_f$) defines the Bragg angle 
$\theta_{\text{\tiny B}}$ and the detection plane is perpendicular to 
$\kb_f$. 
(b) The rocking-curve measurement is equivalent to  
scanning  the (intensity of the) Fourier-space function
$\tilde{\Psi}$ along the direction $\tilde{\kb}_3$ (left). As a result,
$\tilde{\Psi}$ has a dual representation in 
real-space $\tilde{\psi}$  that is non-orthogonal (right). 
(c) From $(\tilde{\kb}_1, \tilde{\kb}_2, \tilde{\kb}_3)$, another 
orthogonal frame $({\kb}_1, {\kb}_2,{\kb}_3)$ can be obtained 
from the rotation of the vector $\tilde{\kb}_3$ so that it alignes 
with the exit-direction ${\kb}_f$ (left). The representation of the 
3D far-field in this representation system is denoted  $\Psi$ 
and corresponds to an orthogonal representation of the exit-field
$\psi$ (right). 
}
\label{Fig:BraggGeometry}
\end{figure}

In this section, we begin by considering the specific geometry involved 
in a simple BCDI measurement implemented in a symmetric two-circle reflection geometry. We assume that the measurement 
is performed in the far-field regime and that the kinematic approximation applies.
Therefore, the exit-field $\tilde{\psi}$ and the diffracted-field 
$\tilde{\Psi}$ are related by a  3D Fourier transformation. 
An important feature of the Bragg geometry is that making small changes in the angular orientation of the sample with respect to the incident beam direction allows the intensity of the diffracted field
$\tilde{\Psi}$ to be measured in 3D. 
However, the direction of Fourier space scanning along the rocking curve is \textit{not} perpendicular to the 
detection plane (as demonstrated in Part I).
Thus, the measurement of $|\tilde{\Psi}|^2$ corresponds to
a non-orthogonal Fourier space frame $(\tilde{\kb}_1, 
\tilde{\kb}_2, \tilde{\kb}_3)$.
The frame corresponding to the specific BCDI geometry we are considering is shown in Fig.~\ref{Fig:BraggGeometry}(a). 
The dual conjugate real-space frame $(\tilde{\eb}_1, \tilde{\eb}_2, \tilde{\eb}_3)$ is also 
non-orthogonal \cite{berenguer13,Yang2019} and is also shown for our case in Fig.~\ref{Fig:BraggGeometry}(b). 
As a result, the 3D inverse Fourier transform (IFT) of $\tilde{\Psi}$ provides a 
rather unintuitive representation of the exit-field\footnote{
We note
that in BCDI $\tilde{\psi}$ is directly interpreted as an image of the
3D sample with lattice-structural sensitivity. In the context of Bragg 
ptychography, $\tilde{\psi}$ represents a partial view of the sample 
as illuminated by the incident beam at a particular location. 
} $\tilde{\psi}$.

To obtain an image of the sample on a natural orthogonal frame, one possible solution is to generate
a pseudo-dataset derived by interpolating the measured pixelated intensity data onto 
a chosen orthogonal frame Fourier space $({\kb}_1, {\kb}_2, {\kb}_3)$. 
Given this type of data as an input, standard BCDI reconstruction algorithms would produce a 3D image of the object expressed in orthogonal coordinates.
In practice, however,  systematic errors are likely to be encountered due to the 
interpolation of low count rate regions of the data that are present in most BCDI measurements.
However, an alternative approach can be imagined by considering the equations in \eqref{FT_2}.
They suggest 
that the evaluated far-field $\tilde{\Psi}$ that is commensurate with the measurement \textit{in average}, and not the data itself, can be
interpolated onto $({\kb}_1, {\kb}_2, {\kb}_3)$, hence providing an 
orthogonal real-space representation.  
Below, we aim to develop a computationally-efficient transformation strategy based on this concept. 
We take advantage of the fact that 
the measurement frames shown in Fig.~\ref{Fig:BraggGeometry}(b-c) are actually very closely related to an orthonormal frame. 
The desired orthonormal frame $({\kb}_1, {\kb}_2, {\kb}_3)$ differs from 
$(\tilde{\kb}_1, \tilde{\kb}_2, \tilde{\kb}_3)$ by a unique rotation of 
$\tilde{\kb}_3$ about the axis defined by $\tilde{\kb}_1$.
Because of this relatively simple rotational relationship, the interpolation between the two frames can be
performed very efficiently, making it suitable to embed within an iterative phase retrieval algorithm.

In the next section, we show how this interpolation can be implemented
in a continuous framework with Fourier transform
operators. Following that, a practical numerical implementation is
 derived. 

\subsection{Continuous derivation with Fourier operators}
\label{SubSecmapping}

Our aim is to use  the \textit{orthogonal} frame $({\eb}_1, {\eb}_2, {\eb}_3)$ depicted 
in Fig.~\ref{Fig:BraggGeometry}-c  as our resultant
 real-space representation frame. 
Starting in this frame makes the Fourier 
pair ${\psi} \rightleftharpoons {\Psi}$ readily available by application of a Fourier transformation. 
However, one cannot incorporate such a transformation into an iterative phase retrieval image reconstruction algorithm because the the far-field $\Psi$ is described in a frame not consistent with the rocking curve measurement, as shown in 
Fig.~\ref{Fig:GeometryOrthoVSnonOrtho}(a,c). 
To solve this problem, we aim 
at providing a direct mapping from the orthogonally represented  $\psi$ 
to the non-orthogonal $\tilde{\Psi}$. 
We start by explicitly deriving equation \eqref{basis_QS} that describes the relationship between $\qb$ 
and $\tilde{\qb}$ for the two-circle symmetric diffraction 
geometry we consider by way of example. From Fig. ~\ref{Fig:BraggGeometry}(b) and (c), we have:
\begin{equation}
  \label{basis_RQ_ortho}
       \Bb_{q} = 
       \left(
       \begin{array}{ccc}
         1    & 0 & 0\\
         0  & 1 & - \sin \theta_{\text{\tiny B}}\\
         0  & 0 &    \cos \theta_{\text{\tiny B}}
       \end{array}
       \right).
\end{equation}
Applying \eqref{FT_2}  then yields:
\begin{equation}
  \label{Psi_non_ortho}
       \begin{array}{rcl}
         {\displaystyle
          \tilde{\Psi}(\tilde{\qb})}   &=& \cos\theta_{\text{\tiny B}} \times
                                       \Psi(\tilde{q}_1 ,\, \tilde{q}_2 - \tilde{q}_3 \sin \theta_{\text{\tiny B}}, \, \tilde{q}_3 \cos \theta_{\text{\tiny B}}).
       \end{array}
\end{equation}
The relation above is precisely the mapping $\psi \rightarrow 
\tilde{\Psi}$ that is needed.  It allows the
``non-orthogonal'' far-field  $\tilde{\Psi}$ to be derived from the ``orthogonal'' representations 
${\psi} \rightleftharpoons {\Psi}$ (see Fig.~\ref{Fig:GeometryOrthoVSnonOrtho}(a,d)). In addition, we have
\begin{equation}
  \label{Omega_psi}
  \begin{array}{rrl}
  \lefteqn{\Psi(\tilde{q}_1 ,\, \tilde{q}_2 - \tilde{q}_3 \sin \theta_{\text{\tiny B}}, \, \tilde{q}_3 \cos \theta_{\text{\tiny B}})} \qquad \qquad &&\\[1em]
    &:=& 
        {\displaystyle
        \int_{\eR^3}
        } 
        \psi(\rb)\,
        e^{-\iota 2\pi [r_1 \tilde{q}_1 \,+\, r_2(\tilde{q}_2 - \tilde{q}_3 \sin \theta_{\text{\tiny B}}) \,+\, r_3 \tilde{q}_3\cos\theta_{\text{\tiny B}}]}
        \,\text{d} \rb\\[1em]
    &=&         
        {\displaystyle
        \int_{\eR^2}
        } 
        \xi(\rb_\bot, \tilde{q}_3\cos \theta_{\text{\tiny B}})\,
        e^{-\iota 2\pi \rb_\bot^T \tilde{\qb}_\bot}
        \,\text{d} \rb_\bot\\[1.5em]
     &=&         
         [\mathcal{F}_\bot \xi] (\tilde{\qb}_\bot, \tilde{q}_3\cos \theta_{\text{\tiny B}}).
    \end{array}
\end{equation}
In these expressions, the coordinates $\tilde{\qb}_\bot :=(\tilde{q}_1, \tilde{q}_2)$ are parallel to the pixel sampling directions in the measurement plane, and are conjugate to $\rb_\bot = (r_1,r_2)$.
Our aim is to define an intermediate function $\xi$ that preserves the behavior and 3D nature of $\psi$, and that can also be acted upon by a 2D Fourier transform $\mathcal{F}_\bot$ that acts on the
first and second spatial coordinates $\rb_\bot$.  This can be done by defining the following: 
\begin{equation}
  \label{xi}
  \xi(\rb_\bot, q_3) := [\mathcal{F}_3 \psi] (\rb_\bot, q_3) \times e^{\iota 2\pi r_2 q_3 \tan \theta_{\text{\tiny B}}} 
\end{equation}
with $\mathcal{F}_3$ the 1D FT operator acting  along the third 
spatial coordinate $r_3$. Finally, \eqref{Psi_non_ortho} and  \eqref{Omega_psi} 
can be combined so that we obtain
\begin{equation}
  \label{final_forward}
  \tilde{\Psi}(\tilde{\qb}) \, = \, \cos\theta_{\text{\tiny B}} \times \left[ \mathcal{F}_\bot \xi \right]
  (\qb_\bot = \tilde{\qb}_\bot, q_3 = \tilde{q}_3 \cos \theta_{\text{\tiny B}}). 
\end{equation}
By examining \eqref{xi}  and \eqref{final_forward}, we see that three distinct steps are required to 
compute the non-orthogonal far field $\tilde{\Psi}$ starting from the real-space orthogonal
representation $\psi$. They can be described as  ($i$)  applying a one-dimensional 
FT, which provides  $\mathcal{F}_3 \psi$; ($ii$) a pointwise 
multiplication, which multiplies to the result of ($i$) with a spatially varying phase term; and ($iii$) 
a FT acting on two of the three axes, which provides $\tilde{\Psi}$.
Considering the problem in terms of these steps indicates a clear path towards numerical implementation. 
Furthermore, the computational burden involved in such a calculation of $\tilde{\Psi}$ from
$\psi$ is similar to a single 3D digital FT.  
This opens up the possibility of incorporating such a computation 
within an iterative reconstruction algorithm without adding 
a significant time penalty and enabling natural frames in both 
real and Fourier spaces to be enforced from the outset. 
This provides distinct benefits, outlined in Section 1, that cannot 
be realized with the approach presented in Part I.

In the context of typical iterative phase retrieval algorithms, we
need to define not only the ``forward'' calculation of $\tilde{\Psi}$ 
from $\psi$ (as in  Relation \eqref{final_forward}), but also a
``backward'' step that computes $\psi$ from $\tilde{\Psi}$. This 
backward step can be deduced from \eqref{Omega_psi}  and 
\eqref{xi} by adopting an intermediate variable $\zeta$ that 
mimics the form of $\xi$:
\begin{equation}
  \label{xi_inv}
  \psi   = {\small \frac{1}{\cos \theta_{\text{\tiny B}}}} \times
  [\mathcal{F}^{-1}_3 \zeta] 
\end{equation}
with  
\begin{equation}
  \label{xi_inv_}
  {\textstyle
  \zeta (\rb_\bot, q_3) := [\mathcal{F}_\bot^{-1} \tilde{\Psi}]
  \big(\tilde{\rb}_\bot=\rb_\bot , \tilde{q}_3 = \frac{q_3}{\cos\theta_{\text{\tiny B}}}\big) \times e^{-\iota 2\pi r_2 q_3 \tan \theta_{\text{\tiny B}}}.
  }
\end{equation}

We note that the transformation between $\psi$ and $\tilde{\Psi}$ involves 
an \textit{implicit} interpolation to appropriately ``shear''  Fourier 
space to conform to the detection geometry as dictated by the experiment. 
The ability to ``shear'' and ``un-shear'' of the Fourier frame is provided by the complex 
exponential terms in the expressions of $\xi$ and $\zeta$,
given by \eqref{xi} and \eqref{xi_inv} respectively.
The form of the exponential terms is, by design, related to a set of linear phase ramp that are frequently encountered in numerically efficient Fourier interpolation methods.
The above derivation leads to several important insights: 
\begin{itemize}
\item If a more complex transformation would be considered, for example a rotation of two vectors in the basis rather than one, the ability to
  perform this interpolation \textit{via} a phase
  ramps would be lost and the 
computational complexity of an alternative approach would increase 
significantly.  
\item  The frame shown in Fig.~\ref{Fig:BraggGeometry}(c) is not 
the only orthogonal frame that one can built from a single
rotation.  Interestingly, another orthogonal frame is obtained  
by the clockwise rotation by $\theta_{\text{\tiny B}}$ of the vector $\tilde{\kb}_2$
about $\tilde{\kb}_1$ which, consequently,  gives an alternative orthogonal 
real-space frame. 
This latter one may be convenient if symmetric Bragg reflexions are involved,
because it often matches a natural ``laboratory frame''.
The derivation of the alternative mapping is straightforward
adaptation of the equations given above, and it is not derived here. 
\item Finally, the simplified  scattering geometry considered in our derivations is limited to 
symmetric Bragg reflections implemented on a two-circle
diffractometer. In a more general Bragg scattering geometry 
obtained with diffractometers with more degrees of freedom, 
the vector $\tilde{\kb}_3$ defining the Fourier space scanning direction of the  RC  will have a non-zero projection along $\kb_1$. 
The extension of the equations
presented in this section to such a situation are presented in Appendix~A.
\end{itemize}

We end by pointing out that the derivations presented above were obtained with operators defined over 
continuous domains. For the practical application to experimental
data, we now consider the discrete sampling in real and Fourier space,
that is inherent to any numerical and experimental implementation.  

\subsection{Implementation with Discrete Fourier Transforms}
\label{numerical_implementation}
Briefly, let us recall first the results obtained
in the previous section. The forward mapping (from $\psi$ to $\tilde{\Psi}$) is 
given by 
\begin{equation}
  \label{psi_Omega_summ}
    \left\{
  \begin{array}{rcl}
    \xi(\rb_\bot, q_3) & =& [\mathcal{F}_3 \psi] (\rb_\bot, q_3) \times e^{\iota 2\pi r_2 q_3 \tan \theta_{\text{\tiny B}}}\\[.5em]
    \tilde{\Psi}(\tilde{\qb})      &=&    \cos \theta_{\text{\tiny B}} \times  [\mathcal{F}_\bot \xi]
                                   (\tilde{\qb}_\bot, \tilde{q}_3\cos \theta_{\text{\tiny B}})
    \end{array}
    \right.
\end{equation}
and the backward mapping (from $\tilde{\Psi}$ to $\psi$) by
\begin{equation}
  \label{Omega_psi_summ}
  \left\{
  \begin{array}{rcl}
    \zeta (\rb_\bot, q_3) &=& [\mathcal{F}_\bot^{-1} \tilde{\Psi}]
    \big(\rb_\bot, \frac{q_3}{\cos\theta_{\text{\tiny B}}} \big) \times e^{-\iota 2\pi r_2 q_3 \tan \theta_{\text{\tiny B}}}\\[.5em]
    \psi(\rb)  &=& \frac{1}{\cos \theta_{\text{\tiny B}}} \times [\mathcal{F}^{-1}_3 \zeta ] (\rb)     
    \end{array}
    \right.
\end{equation}
where $\xi$ and $\zeta$ are intermediate functions designed to 
enable convenient separation of the 3D Fourier transformation 
integral into sequential operations involving 2D and 1D Fourier transformations.

However, numerical evaluation of these quantities requires discrete
sampling of the real and Fourier-space domains. The 
Fourier-space mesh is defined by
\begin{equation}
  \label{sampling_QS}
      {\qb} \in \{\Lambdab_{{q}} {\mb} \} 
      \quad \text{where} 
      \quad 
     \Lambdab_{{q}} = 
      \left(
      \begin{array}{@{\kern0pt}c@{\kern0pt}c@{\kern0pt}c@{\kern0pt}}
        \delta_{{q}_1} &&\\
        & \delta_{{q}_2}&\\
        && \delta_{{q}_3}
        \end{array}
        \right).
\end{equation}
In this expression, ${\mb} := [{m}_1~~{m}_2~~{m}_3]^T \in \eZ^3$ is the Fourier space
pixel index, and  ${\delta}_{q_1}$, ${\delta}_{q_2}$, ${\delta}_{q_3}$ are the 
Fourier-space sampling-rates along $\kb_1$, $\kb_2$ and $\kb_3$,
respectively. 
Because we chose the orthogonal
and the non-orthogonal frames such that $\kb_1 = \tilde{\kb}_1$ 
and $\kb_2 = \tilde{\kb}_2$, both frames differ only because
$\kb_3 \neq \tilde{\kb}_3$, as seen in Fig.~\ref{Fig:BraggGeometry}. 
As a consequence, ${\delta}_{q_1}$ and ${\delta}_{q_2}$ can be derived in a straightforward way, as would be done for a 
standard transmission-geometry CDI experiment: 
\[
\delta_{q_1} \equiv \delta_{q_2} = \frac{1}{\lambda} \times \frac{p}{D},
\]
where $\lambda$ is the x-ray wavelength, $D$ is the sample-to-detector distance,
and $p$ is the pixel pitch of the camera. In the BCDI geometry, the sampling rate $\delta_{q_3}$ acting along
$\kb_3$ must be considered more carefully. The Fourier sampling 
increment expressed on the orthogonal basis is arrived upon by 
projecting the sampling increment $\delta_{\tilde{q}_3}$ from the 
non-orthogonal frame:
\begin{equation}
  \label{sampling_QS_x}
  \delta_{q_3} = \delta_{\tilde{q}_3} \cos \theta_{\text{\tiny B}}.
\end{equation}
Figure~\ref{Fig:BraggGeometry} shows this relationship geometrically. 
We note that the expression for $\delta_{q_3}$ is not given in terms
of more fundamental experimental parameters because it is dependent on 
the choice of angular increment that the experimenter implements in
the rocking curve measurement. A detailed discussion of how to derive 
$\delta_{q_3}$ in terms of these experimental parameters is given in 
Appendix~A, and in Part I for the case of a more general diffraction geometry.

The sampling in orthogonal real-space that corresponds to a rocking curve scan in Fourier space can be expressed in terms that mirror 
 \eqref{sampling_QS} above, but certain subtleties arise that must not be overlooked. 
We start by observing that the sampling in real space can, as above, be expressed by the following:
\begin{equation}
  \label{sampling_RS}
      \rb \in \{\Lambdab_{{r}} \nb \}
      \quad 
      \text{where} 
      \quad
      \Lambdab_{r} = 
      \left(
      \begin{array}{@{\kern0pt}c@{\kern0pt}c@{\kern0pt}c@{\kern0pt}}
        \delta_{{r}_1} &&\\
        & \delta_{{r}_2}&\\
        && \delta_{{r}_3}
        \end{array}
        \right).
\end{equation}
Here, $\nb := [{n}_1~~{n}_2~~{n}_3]^T\in \eZ^3$ is the real space 
sample index, and  $\delta_{{r}_1}$, $\delta_{{r}_2}$, 
$\delta_{{r}_3}$ are the real space sampling rates
along $\eb_1$, $\eb_2$ and $\eb_3$, respectively.
However, an important difference arises between Fourier and real space sampling when we consider the 
 magnitudes of the sampling increments. 
Whereas in Fourier space, sampling along the first two bases ($\delta_{q_1}$, $\delta_{q_2}$) are equivalent owing to the square pixelation of the detector, 
this is not the case in orthogonal real space. 
A consequence of mapping the naturally non-orthogonal Fourier space volume of a rocking curve to orthogonal real space is that the real space sampling increments couple to the geometry of the measurement in a non-intuitive manner. 

\begin{figure}
\centering
\includegraphics[width=1\textwidth]{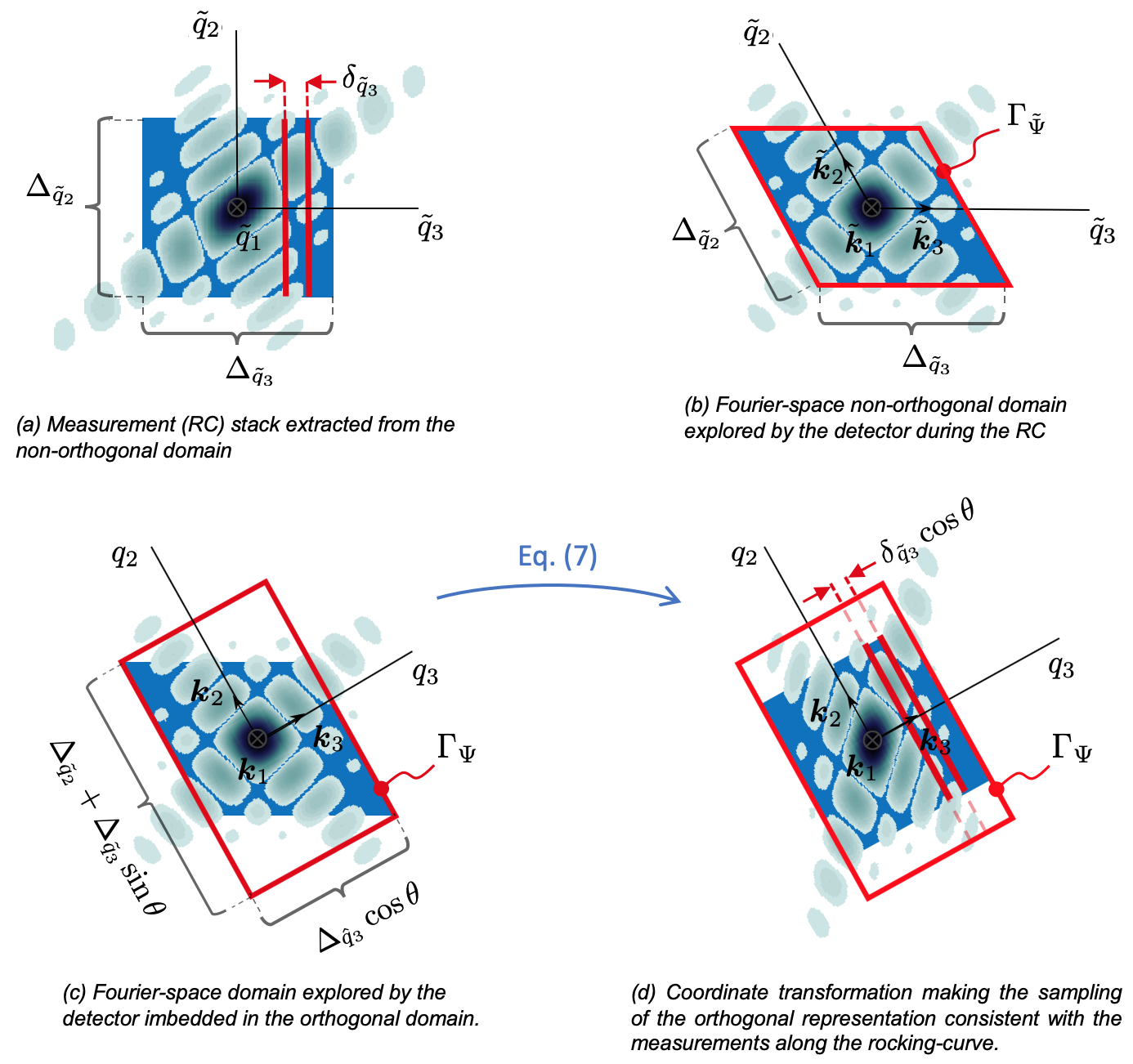}
\caption{Non-orthogonal and orthogonal representations of the 
diffracted far-field in the Bragg geometry. 
(a) The 3D stack of intensity measurements obtained during the RC, as
stored in a 3D array, (b) is produced by the regular sampling  of the 
far-field $\tilde{\Psi}$ performed within the non-orthogonal 
measurement domain imposed by the detector geometry and delimited 
by the shaded area $\Gamma_{\tilde{\Psi}}$. 
(c) In an orthogonal basis, the far-field $\Psi$ computed
in a domain $\Gamma_{\Psi}$ can de defined to contain the non-orthogonal 
domain $\Gamma_{\tilde{\Psi}}$. In addition, after the coordinate
transformation given in \eqref{Psi_non_ortho}, $\Psi$ provides a computed 
representation of the far-field whose sampling is totally consistent with 
the data (RC) stack (d). Because the 
non-orthogonal measurement domain $\Gamma_{\tilde{\Psi}}$ 
shown in (d) is smaller than $\Gamma_{\Psi}$, the computed
representation of $\tilde{\Psi}$ in $\Gamma_{\Psi}$ is not totally 
constrained by the intensity measurements shown in (b). For 
the sake of the demonstration, a cubic sample has been numerically
designed, with its Bragg vector pointing towards one of its edges.
Note the agreement between (a) and (d).
}
\label{Fig:GeometryOrthoVSnonOrtho}
\end{figure}

To aid in deriving the real space sampling increments, it is useful to consider the geometric constructions presented in Fig.~\ref{Fig:GeometryOrthoVSnonOrtho}. 
In this Figure, a 2D cut through synthetic BCDI data are shown from a
cubic sample. In the Fourier domain shown in
Fig.~\ref{Fig:GeometryOrthoVSnonOrtho}, the structure of the cube is
encoded as a 2D sinc function diffraction intensity pattern. In each panel, the sinc function
fringes are recognizable, but manifest themselves in different ways 
by manipulating the basis vectors of the Fourier volume interrogated 
by the rocking curve. Manipulating these bases in a specific manner 
will enable us to define the real space sampling increments needed in 
\eqref{sampling_RS}.

In Figure~\ref{Fig:GeometryOrthoVSnonOrtho}(a), the 2D sinc function 
is shown in the measurement frame, equivalent to the 3D ``data stack'' 
from the rocking curve. The fringes of the 2D sinc function in this
frame are clearly not perpendicular, and the fringe oscillation period 
along the two primary axes of the cube are not the same. This
distortion is due to the shear introduced by the inherently
non-orthogonal nature of any rocking curve
measurement. Figure~\ref{Fig:GeometryOrthoVSnonOrtho}(b) 
maps the data stack to a frame where the basis vectors $\tilde{\kb}_2$ and $\tilde{\kb}_3$ are non-orthogonal, 
to the degree dictated by the geometry of the experiment. In this 
frame, the fringes of the 2D sinc function are perpendicular 
and the oscillation period along both fringe directions is the same. 
In this frame, we recognize the basic symmetry and orientation 
of the cube used to generate this data. 
What is now sheared is the volume of Fourier space surveyed 
by the rocking curve. This representation is therefore useful, 
but it is built upon on a non-orthogonal Fourier basis that is 
not amenable to a description of orthogonal real space sampling of the object. 

To alleviate this, we cast the rocking curve volume onto an orthogonal 
Fourier basis, as shown in Fig.~\ref{Fig:GeometryOrthoVSnonOrtho}(c). 
In this frame, capturing all of the information contained in the rocking curve 
requires a rectangular area along the orthogonal basis vectors
$\kb_2$, $\kb_3$. 
This rectangular Fourier space domain, 
denoted as $\Gamma_{\Psi}$, will  define the Fourier space window used in an iterative phase retrieval algorithm that enforces an orthogonal view of the sample. 
In this context, we should account for the fact that
\eqref{psi_Omega_summ} and \eqref{Omega_psi_summ} will be computed 
by discrete Fourier transforms (DFT).
In that case, our construction must adhere to the following constraint, which stipulates that the 
sampling rates in real and Fourier space are related:  
\begin{equation}
 \label{TFD1}
 {\rb}^T {\qb} = \nb^T \Lambdab_{{r}}
 \Lambdab_{q} \mb = \nb^T \Lambdab \mb
 \quad \text{with} 
 \quad 
 \Lambdab = 
     \left(
     \begin{array}{@{\kern0pt}c@{\kern0pt}c@{\kern0pt}c@{\kern0pt}}
       N_1^{-1} &&\\
       & N_2^{-1}&\\
       && N_3^{-1}
       \end{array}
       \right)
\end{equation}
In this equation, $N_1$, $N_2$ and $N_3$ are the number of sample points 
in real space and in Fourier space, corresponding to the directions
 ${\eb}_1$ and ${\kb}_1$, ${\eb}_2$ and ${\kb}_2$, and ${\eb}_3$ 
and ${\kb}_3$, respectively. 
Thus a single 3D array of pixels should be defined that will be iteratively transformed from real to Fourier space and back over the course of phase retrieval. 
However, in imagining such an approach, a complication arises because the extent of $\Gamma_{\Psi}$ along the $\qb_2$ direction, 
given by $(\Delta_{\tilde{q}_2} + \Delta_{\tilde{q}_3}\sin \theta_{\text{\tiny B}})$ as shown visually in Fig.~\ref{Fig:GeometryOrthoVSnonOrtho}(c), is not necessarily divisible by $\delta_{q_2}$ leading to a situation where $N_2$ is not an integer number of pixels. 
This can be addressed by simply rounding up so that the integer pixelation is enforced. 
Thus, we have $N_2 = {\ceil {(\Delta_{\tilde{q}_2} + \Delta_{\tilde{q}_3}\sin \theta_{\text{\tiny B}})/\delta_{q_2}}},  $ where $\ceil{\;}$ is a round-up operation.  
Numerically, this creates a domain that is slightly larger than $\Gamma_\Psi$, but that retains the important property that it still totally encapsulates the RC measurement volume. 
We note that a similar rounding consideration should also be made for $N_1$ if one considers more general non-symmetric Bragg CDI measurements, but this is not necessary for the simpler geometry considered here. 

Having arrived at a DFT-compatible description of $\Gamma_\Psi$, we can readily derive the corresponding real space increments of the 3D pixel array. 
This is done by using the extent of the
Fourier domain $\Gamma_\Psi$ to determine the magnitudes of the 
sampling rates along the three orthogonal directions in real space, which will
generally all be different. In each of the orthogonal directions, 
we have:
\begin{equation}
  \label{sampling_RS_x}
  \delta_{r_3}  = \frac{1}{\Delta_{\tilde{q}_3}\cos \theta_{\text{\tiny B}}}, 
\end{equation}
\begin{equation}
  \label{sampling_RS_yz}
  \delta_{r_1}  = \frac{1}{\Delta_{\tilde{q}_1}} 
  \quad \text{and} \quad 
    \delta_{r_2}  = \frac{1}{\delta_{q_2}} \times \frac{1}{\ceil {(\Delta_{\tilde{q}_2} + \Delta_{\tilde{q}_3}\sin \theta_{\text{\tiny B}})/\delta_{q_2}}}.  
\end{equation}
In these relations, $\Delta_{\tilde{q}_3}$  is the extent of $\Gamma_\Psi$ along
the rocking curve direction $\tilde{\kb}_3$, and
$\Delta_{\tilde{q}_1}$ and $\Delta_{\tilde{q}_2}$ are the Fourier extents of the detector 
along  $\tilde{\kb}_1$ ($\equiv\kb_1$) and $\tilde{\kb}_2$
($\equiv\kb_2$), respectively.
The case of $\delta_{r_2}$ is particularly interesting because it is derived from 
an edge length of $\Gamma_\Psi$ that exceeds the Fourier space window 
subtended by the pixels of the detector. 
It is related to the 
\emph{total} extent along $\kb_2$ interrogated by the entire
measurement within the orthogonal frame. Similarly, $\delta_{r_3}$ is 
determined by the height of the parallelogram rather than the rocking 
curve extent in the original measurement frame. 
These derivations build a stronger connection with the details of the physical experiment and are not readily apparent when only considering the
``sheared'' measurement domain $\Gamma_{\tilde{\Psi}}$ shown in 
Fig.~\ref{Fig:GeometryOrthoVSnonOrtho}(b).

A final consideration in our construction is that
the far-field diffraction pattern $\Psi$ shown in 
Fig.~\ref{Fig:GeometryOrthoVSnonOrtho}(c), though representing the cubic symmetry we anticipate, is not actually sampled at Fourier space sample points corresponding to the data measured with a rocking curve.
As a result, it would not be possible, for example, to impose a modulus constraint of intensity measurements using this particular Fourier representation. 
A final transformation, summarized in 
Eq.~\eqref{Psi_non_ortho}, must be applied so that the far field corresponding to an orthogonal real space frame is sampled at ``sheared'' points corresponding to the measurement. 
As shown in 
Fig.~\ref{Fig:GeometryOrthoVSnonOrtho}(d), the resulting geometric 
transformation yields a description of the far field that closely matches Fig.~\ref{Fig:GeometryOrthoVSnonOrtho}(a), in that 
the fringes of the sinc function are no longer perpendicular and they are unevenly 
spaced. 
This may seem like a circular exercise, as we have
once again lost the direct connection to the underlying symmetry 
of the cube used to synthesize the data. 
However,  the geometry of the domain $\Gamma_\Psi$ and the transformation performed with
Eq.~\eqref{Psi_non_ortho} in the orthogonal Fourier frame together retain, by construction, all the components needed to map back to the orthogonal real space representation of the scattering object.
Namely, 
$i$) we can easily deduce $\Gamma_{\Psi}$ from the RC parameters;
$ii$) $\Gamma_{\Psi}$ determines the numerical array size appropriate to capture all the Fourier information of the RC in an orthogonal frame;
$iii$) we can use $\Gamma_{\Psi}$ to determine the corresponding orthogonal real space sampling intervals 
and $iv$) the nature of the phase-ramp terms associated with the
intermediate functions $\xi$ and $\zeta$, which stem from Eq.~\eqref{Psi_non_ortho}, are such that they exactly encode 
the degree of shearing of the RC domain within $\Gamma_\Psi$, thus linking Fig.~\ref{Fig:GeometryOrthoVSnonOrtho}(d) to (c).

With the above construction, the numerical evaluation of $\tilde{\Psi}$ given in \eqref{psi_Omega_summ} 
is now discussed, starting  by defining the following 3D arrays:
\begin{equation}
  \label{numericalQuantities}
  \begin{array}{rcl}
    \psib & \equiv & \{\psi(\Lambdab_{r}  \nb) \} \in \eC^{N_2 \times N_1 \times N_3}\\[.5em]    
    \tilde{\Psib} & \equiv & \{\tilde{\Psi}(\Lambdab_{q}  \mb) \} \in \eC^{N_2 \times N_1 \times N_3}\\[.5em]
    \mub & \equiv & \{  e^{\iota 2\pi R \times m_3 n_2}, \, \forall n_1 \} \in \eC^{N_2 \times N_1 \times N_3}
    \end{array}
\end{equation}
where $R = \delta_{r_2} \delta_{\tilde{q}_3} \sin \theta_{\text{\tiny B}}$. 
Numerical implementation of \eqref{psi_Omega_summ} would then take the following form:
\begin{equation}
  \label{psi_num}
      \tilde{\Psib} \,= \, \textbf{DFT}_\bot\big(\textbf{DFT}_3 (\psib) \,
     \odot \, \mub \big) \times \delta_{r} \cos \theta_{\text{\tiny B}}.  
\end{equation}
Here, $\textbf{DFT}_3$ is a 1D-DFT operator that acts along 
the third dimension of a 3D array, $\textbf{DFT}_\bot$ is a 
2D-DFT operator acting along the two other array dimensions.
The $\odot$ symbol represents the component-wise multiplication between matrices,
and $\delta_{r}\equiv \delta_{r_1}\delta_{r_2}\delta_{r_3}$ is the
real-space voxel volume. The conjugate operation to \eqref{psi_num}, 
derived from from  \eqref{Omega_psi_summ}, would then read:
\begin{equation}
  \label{omega_num}
      \psib \, = \, \textbf{IDFT}_3\big(\textbf{IDFT}_\bot (\tilde{\Psib}) \,
     \odot \, \mub^* \big) \times \frac{1}{\delta_{r}\cos \theta_{\text{\tiny B}}}
\end{equation}
where $\textbf{IDFT}_3$ and $\textbf{IDFT}_\bot$ are the inverse of
the  DFT operators introduced above, and  '$*$' the symbol denoting
a component-wise complex-conjugate operation.  

Because these transformations will be performed in the context of phase retrieval, we also note that Equations
\eqref{psi_num} and \eqref{omega_num} should adhere to the
Shannon-Nyquist sampling theorem.
In this work, we assume that the oversampling criteria that come about from this theorem are more than satisfied such that
any aliasing induced by the spatial sampling is negligible and
that the one period of the DFT is essentially identical to the original
continuous FT.

Equations \eqref{psi_num} and \eqref{omega_num} provide the means by which to design a phase retrieval algorithm in which the sample frame is orthogonal. 
The 3D diffracted field $\tilde{\Psib}$ computed with
\eqref{psi_num} is consistent with the RC intensity measurements. Thus, this 
equation  can be used within a 3D BCDI phase-retrieval strategy to 
enforce a ``data-constraint'' or ``modulus constraint'' for each iteration that is mapped to orthogonal real space \textit{via} 
\eqref{omega_num}. 
It should also be noted that, in imposing such a modulus constraint, some sample points of the numerical array
will not be constrained by intensity measurements of the RC because of our construction of $\Gamma_\Psi$. 
Generally, this is not a concern because phase retrieval algorithms can be designed to handle a set of
 ``missing'' or ``floating'' data points \cite{Nishino03,Marchesini03,berenguer13}.

\begin{figure}
\centering
\includegraphics[width=.7\textwidth]{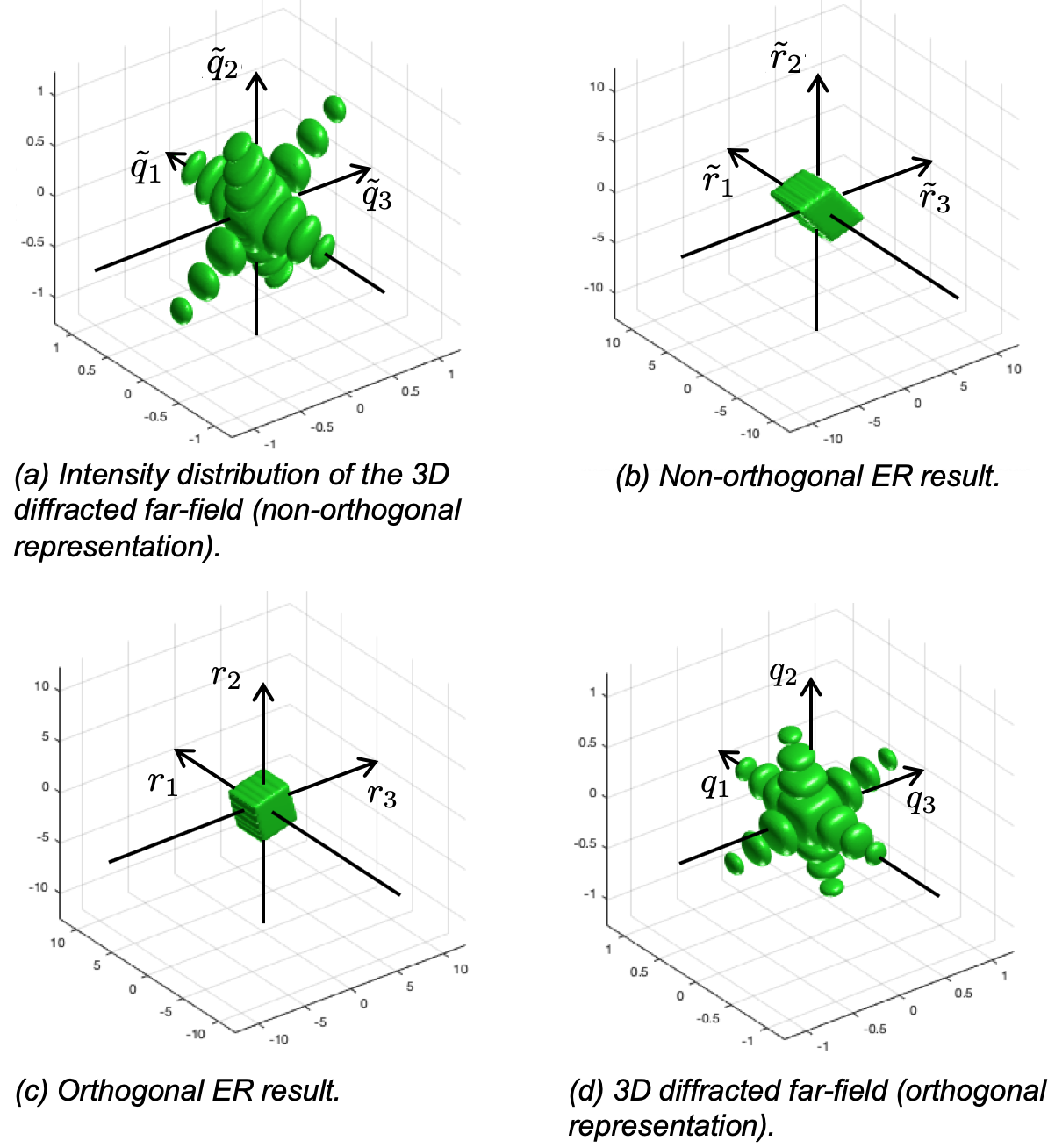}
\caption{Reconstruction of a uniform, real-valued cubic sample from
  the intensity of the 3D diffracted field measured along the RC. (a)
  the stack of noise-free intensity patterns collected along the RC is identical
  to the intensity of the non-orthogonal representation of the 3D 
  diffracted-field $\tilde{\Psi}$. (b) From these noise-free measurements, the 
sample estimate provided with the standard ER iteration is provided in
a non-orthogonal frame, hence producing geometrical distortions 
of the cubic sample. On the contrary, the modified ER algorithm
(given in Appendix~B) provides an orthogonal representation for the
sample estimate (c) and its 3D diffracted far-field (d). The results
shown in (b) and (c) are obtained with 100 iterations of ER with a perfect
knowledge of the support of the sample, either in its non-orthogonal 
representation  for (b) or in its orthogonal representation for
(c). The mesh size is $N_1\times N_2\times N_3= 250^3$ and the 
computational-time per iteration with a regular laptop is 0.96s 
for (b) and 1.3s for (c).}
\label{Fig:ReconstructionSquare3DFT}
\end{figure}

We demonstrate the effectiveness of utilizing the specialized Discrete Fourier transformations proposed in this section in an actual phase retrieval problem
 in Fig.~\ref{Fig:ReconstructionSquare3DFT}. 
 Featured is a
reconstruction result obtained by employing the popular and well known iterative Error-Reduction (ER) phase retrieval strategy for 3D BCDI adapted with the above transformations.
To aid in implementing this approach, we
provide in Appendix~B the \textsc{matlab} code that
implements this strategy for ER.

\section{Alternative BCDI mappings \textit{via} projections and back-projections}
\label{ProjBAckProj}
As we show below, continuous FT/IFT operators can also be interpreted as 
\textit{projection} and \textit{back-projection} operations, consistent with the Fourier slice projection theorem.
This interpretation provides the means by which to devise a second strategy to link either $\tilde{\psi}$ or
$\psi$ to the ``measurement space'' $\tilde{\Psi}$ with even greater flexibility.
This flexibility offers the possibility to account for more complex physical models of the system. 
For example, the requirement that a rocking curve be measured with evenly spaced angular increments can be relaxed with this approach. 
We note that the flexibility of this back-projection approach has been employed in several recent works as a means to solve specific problems within BCDI and Bragg ptychography that required the incorporation of more physically realistic models of the experiment to enable new types of measurements. 
Examples include, determination of the angular errors of a BCDI rocking curve~\cite{calvo-almazan19}, accounting for Fourier space scaling in a BCDI energy-scanning measurements~\cite{cha16}, dealing with raster grid misregistry in 3D Bragg ptychography~\cite{hill18}, and enabling 3D reconstructions from 2D Bragg ptychography data measured at a fixed angle~\cite{hruszkewycz17}. 
Though the approach has been used effectively in different contexts in the literature, a consolidated mathematical derivation and roadmap for numerical implementation is lacking. 
Thus, the goal of this section is to derive the relevant relationships 
and strategies for their numerical evaluation \textit{via} DFT/IDFT to facilitate the adoption of this concept towards yet more BCDI applications.

\subsection{Mappings between non-orthogonal representations}
\label{nonOrtgogonalProjBProj}
A general property of an $n$-dimensional FT operator, 
regardless of the representation frame, is that it can be implemented as a series of
lower dimensional FTs, a fact we will take advantage of in this section. 
We start by considering the Fourier pair in the ``non-orthogonal''
representation $\tilde{\psi} \rightleftharpoons \tilde{\Psi}$: 
\begin{equation}
  \label{sep_FT_1}
  \tilde{\Psi} \, = \, \Fc \tilde{\psi} \,=\, \Fc_\bot \Fc_3 \tilde{\psi}.
\end{equation}
We next define a notation to designate a single 2D slice in the measurement frame that corresponds to a given angle that might be interrogated by a rocking curve. 
This could be a Fourier space coordinate that is one of a series of regular rocking curve intervals, but it need not be. The notation we adopt is that $\tilde{q}_{3;0}$ represents an arbitrary point along the $\tilde{q}_3$ measurement axis. At the point $\tilde{q}_{3;0}$, we expect that a 2D cut of the far field pattern could be expressed as:
\[
  \begin{array}{rcl}
    \tilde{\Psi}_{\bot;\tilde{q}_{3;0}}  (\tilde{\qb}_\bot)
    &:=& 
         \tilde{\Psi}  (\tilde{\qb}_\bot, \tilde{q}_{3} = \tilde{q}_{3;0})\\[1em]
    &=&
        {\displaystyle
        \int_{\eR^3}
        } 
        \tilde{\psi}(\tilde{\rb})\,
        e^{-\iota 2\pi [\tilde{r}_1 \tilde{q}_1 \,+\, \tilde{r}_2 \tilde{q}_2  \,+\, \tilde{r}_3 \tilde{q}_{3;0}]}
        \,\text{d} \tilde{\rb}.
    \end{array}
\] 
Alternatively, the integral above can be re-arranged so that this 2D
cut $\tilde{\Psi}_{\bot;  \tilde{q}_{3;0}}$ is obtained \textit{via} the
2D Fourier transform of a specific 2D projection 
of the diffracting object:
\begin{equation}
  \label{sep_FT_2}
  \begin{array}{rcl}
  \tilde{\Psi}_{\bot;\tilde{q}_{3;0}} 
    &=& 
         \mathcal{F}_\bot \tilde{\psi}_{\bot;  \tilde{q}_{3;0}}. 
    \end{array}
\end{equation}
In this equation, we introduce a new 2D function $\tilde{\psi}_{\bot;
  \tilde{q}_{3;0}}$ that represents  the appropriate projection of the
3D object needed to produce a specific cut along the rocking curve 
of the measurement-frame diffraction pattern. 
It is given by:
\begin{equation}
  \label{sep_FT_3}
    \tilde{\psi}_{\bot;\tilde{q}_{3;0}}(\tilde{\rb}_\bot)     :=  \int_{\eR}
    \tilde{\psi}(\tilde{\rb}) \times e^{-\iota2\pi \tilde{r}_3 \tilde{q}_{3;0}}       \,\text{d} \tilde{r}_3.
\end{equation}
Central to this projection is the fact that the 3D object is to be multiplied by a complex exponential that imparts an oscillating, spatially modulated phase on the object. 
The period of this phase oscillation encodes the real-space spatial frequency along $\tilde{r}_3$ that corresponds to the desired cut of the Fourier-space diffraction pattern at $\tilde{q}_{3;0}$.
Thus, \eqref{sep_FT_2}  provides a forward mapping 
$\tilde{\psi} \rightarrow \tilde{\Psi}$ in which a full 3D Fourier-space description of $\tilde{\Psi}$ can be built up slice-by-slice by invoking a series of 2D Fourier transformations upon specific projections of the object that include appropriate phase modulation. 

Conversely, the complimentary backwards transformation
$\tilde{\Psi}\rightarrow \tilde{\psi}$ can also be derived by a similar construction. 
We first consider the result of a 3D inverse Fourier transform acting upon a single 2D slice of the far field diffraction. 
Because only a 2D slice of the 3D far field is input in this transformation, we do not expect the resulting real-space description of the object to be complete. 
Thus, we designate the quantity $\tilde{\psi}_{\tilde{q}_{3;0}}(\tilde{\rb})$ to represent the 3D description of the object consistent with the information contained in the   $\tilde{\Psi}_{\bot;  \tilde{q}_{3;0}}$ slice:
\begin{equation}
  \label{backprop}
  \begin{array}{rcl}
  \tilde{\psi}_{\tilde{q}_{3;0}}(\tilde{\rb}) 
    &:=& 
        {\displaystyle
        \int_{\eR^3}
        }
         \tilde{\Psi}_{\bot;  \tilde{q}_{3;0}} (\tilde{\qb}_\bot) 
         \times \delta(\tilde{q}_{3}  - \tilde{q}_{3;0})
        \,
        e^{\iota 2\pi \tilde{\rb}^T \tilde{\qb}}
        \,\text{d} \tilde{\qb}\\[1em]
     &=&         
        e^{\iota 2\pi \tilde{r}_3 \tilde{q}_{3;0}}  \times \big[\Fc_\bot^{-1}
         \tilde{\Psi}_{\bot;  \tilde{q}_{3;0}}\big] (\tilde{\rb}_\bot) \\[1em]                
    &=&
        e^{\iota 2\pi \tilde{r}_3 \tilde{q}_{3;0}}  \times  \tilde{\psi}_{\bot;\tilde{q}_{3;0}} (\tilde{\rb}_\bot),                    	     
    \end{array}
\end{equation}
where $\delta$ is the Dirac distribution. 
To obtain a complete description of the object $\tilde{\psi}({\tilde{\rb}})$, we must integrate over all of the possible $\tilde{\psi}_{\tilde{q}_{3;0}}(\tilde{\rb}) $ that correspond to the continuum of available slices of a far field diffraction pattern:
\begin{equation}
  \label{sum_psi_nonortho}
  \tilde{\psi} (\tilde{\rb}) \, = \,\int_\eR \tilde{\psi}_{\tilde{q}_{3;0}}(\tilde{\rb}) \,\text{d}  \tilde{q}_{3;0}. 
\end{equation}

We note that Eq.~\eqref{backprop} represents a back-projection that acts as a complimentary operation to the projection in Eq.~\eqref{sep_FT_3}.
As in the case of the projection, the back-projection takes on a specific nature with regard to the phases of the real-space description of the object. 
We find that the 3D quantity $\tilde{\psi}_{\tilde{q}_{3;0}}$ is composed of the 2D field $\tilde{\psi}_{\bot;\tilde{q}_{3;0}} $ that is self-similar along the direction $\tilde{r}_3$, but for a spatially modulated phase. 
The two phase modulation terms associated with projection in the forward transformation and with back-projection in the backward transformation are simply related complex conjugates because they factor out of the forward and inverse Fourier integrals respectively. 
This particular construction of Fourier transformations emphasizes the ``slice-by-slice'' nature of the measurement rather than treating the rocking curve as an explicitly 3D interrogation of Fourier space. 
This emphasis provides substantially more flexibility. 
For example, with the projection / back-projection framework, appropriate transformations can easily be designed to account for data measured at arbitrary or irregular angles along a rocking curve.
Additionally, with this mapping from a non-orthogonal description of the far field $\tilde{\Psi}$ to an orthogonal real-space description of the sample $\psi$ is also possible, as we show below.

\subsection{Linking orthogonal real-space and non-orthogonal measurement space  via projections and back-projections}
\label{OrtgogonalProjBProj}

Projection and back-projection also provide a convenient way to compute the ``non-orthogonal'' 
representation $\tilde{\Psi}$ from the real-space ``orthogonal''
representation $\psi$, and \textit{vice versa}. 
We first consider the reciprocal-space displacement that is needed 
from the center of the diffraction pattern to extract our slice of 
interest $\tilde{\Psi}_{\bot;  \tilde{q}_{3;0}}$. 
This displacement vector, or shift vector, is to be determined in the orthogonal $\qb$ frame. 
For the case of symmetric two-circle diffraction considered here, the three coordinates of the shift vector as expressed in the orthogonal frame are:
\begin{equation}
  \label{shift}
  \begin{array}{rcl}
    \Deltab_{\tilde{q}_{3;0}}
    &=& 
\tilde{q}_{3;0} \cos\theta_{\text{\tiny B}} \times 
	  \left[
    0~~-\tan \theta_{\text{\tiny B}}~~1
	  \right]^T
        \\[1em]
    &=&
	  \left[
    0~~-\tilde{q}_{3;0} \sin\theta_{\text{\tiny B}}~~\tilde{q}_{3;0} \cos\theta_{\text{\tiny B}}
	  \right]^T.
    \end{array}
\end{equation}
The geometry of this shift vector is consistent with the means of deriving $\Gamma_\Psi$ earlier, as shown in Fig.~2.

This shift vector can now be used in combination with Eq.~\eqref{Psi_non_ortho} in order to map the Fourier space measurement frame to an orthogonal one: 
\begin{equation}
  \label{TCPGen}
  \tilde{\Psi}_{\bot;  \tilde{q}_{3;0}} (\tilde{\qb}_\bot) \,=\, \cos
  \theta_{\text{\tiny B}} \times \Psi(\qb +
  \Deltab_{\tilde{q}_{3;0}})|_{\qb_\bot = \tilde{\qb}_\bot, q_3 = 0}. 
\end{equation}
As above, we note that a given slice of the far field pattern $ \tilde{\Psi}_{\bot;  \tilde{q}_{3;0}} (\tilde{\qb}_\bot)$ is related to a Fourier transform of a projection of the object subject to specific phase modulation. 
This also holds true when describing the object in an orthogonal frame, so long as the appropriate shift vector (expressed in the dual orthogonal Fourier frame) is used to determine the phase modulation term. So, we define the following quantity:
\begin{equation}
  \label{mod_psi1}
 \psi_{\bot; \Deltab_{\tilde{q}_{3;0}}}  (\rb_\bot)  \,:=\, \cos
 \theta_{\text{\tiny B}} \times \int_{\eR} \psi(\rb) \times e^{-\iota 2\pi \rb^T \Deltab_{\tilde{q}_{3;0}}} \,\text{d}r_3,
\end{equation}
such that:
\begin{equation}
  \label{ProjGen}
  \tilde{\Psi}_{\bot;  \tilde{q}_{3;0}} (\tilde{\qb}_\bot)    \, = \, \big[ \Fc_\bot \psi_{\bot; \Deltab_{\tilde{q}_{3;0}}} \big] (\qb_\bot= \tilde{\qb}_\bot). 
\end{equation}
There is a parallel to be drawn between $\psi_{\bot;
  \Deltab_{\tilde{q}_{3;0}}}$ and the intermediate functions $\xi$ and
$\zeta$ from Section~\ref{SubSecmapping}. 
These functions are constructed in a way to rectify the difference between orthogonal and non-orthogonal conjugate frames by utilizing spatially varying phase terms.
They are also constructed such that the details of accounting for the skewed RC measurement axis are separated from the $\qb_\bot$ and $\rb_\bot$ coordinates such that a 2D Fourier transformation can be 
deployed\footnote{
Because we have $\tilde{\rb} = \Bb^{-1} \rb$ with $\Bb^{-1}=\Bb_q^T$, 
we can show from \eqref{basis_RQ_ortho} that $\tilde{\rb}_\bot =
\rb_\bot$. This result means that the ``projected exit-fields''
$\psi_{\bot;  \Deltab_{\tilde{q}_{3;0}}} $ and $\tilde{\psi}_{\tilde{q}_{3;0}}$ 
given in \eqref{mod_psi1} and \eqref{sep_FT_3}, respectively,  are indeed 
identical mathematical expressions.}.

We can also derive the series of backward operations that constructs $\psi$ from $\tilde{\Psi}$.
The derivation is based on Eq.~\eqref{sum_psi_nonortho} used in combination with the fact  
(from Sec.~\ref{secQSCT}) that $\psi(\rb) 
\equiv \tilde{\psi}(\tilde{\rb} =\Bb^{-1}\rb)$
with $\Bb^{-1} =\Bb_q^T$. 
We start by re-casting the result of back-projection from a single slice of the diffraction field from the measurement frame into the orthogonal sample frame:
\begin{equation}
  \label{backproportho}
  \begin{array}{rcl}
    \psi_{\tilde{q}_{3;0}}(\rb) 
    &:=& \tilde{\psi}_{\tilde{q}_{3;0}}(\tilde{\rb} = \Bb^{-1}\rb) \\[.5em]
     &=&   e^{\iota 2\pi \rb^T \Deltab_{\tilde{q}_{3;0}}}  \times       
        \big[\Fc_\bot^{-1} \tilde{\Psi}_{\bot;  \tilde{q}_{3;0}}\big]
         (\rb_\bot).        
    \end{array}
\end{equation}
As before, the information regarding the degree of shear of the rocking curve is included in the complex exponential term. 
This leads us again to an expression in which the orthogonal description of the sample can be obtained by integrating over all possible diffraction pattern slices available in the measurement frame:
\begin{equation}
  \label{sum_psi_ortho}
      \psi (\rb) \, := \, \tilde{\psi}(\Bb^{-1}\rb) \,=\, 
    \int_{\eR}    \psi_{\tilde{q}_{3;0}}(\rb) \,\text{d}  \tilde{q}_{3;0}. 
\end{equation}

Equations \eqref{ProjGen}  and \eqref{sum_psi_ortho}  are the main results of this sub-section.
These equations links any ``slice'' in
$\tilde{\Psi}$ with the orthogonal real-space representation 
$\psi$. 
Thus, these two equations can be utilized within a phase-retrieval 
algorithm to directly update an orthogonal real-space representation regardless of the order or regularity with which the slices were measured experimentally.

\begin{figure}
\centering
\includegraphics[width=.9\textwidth]{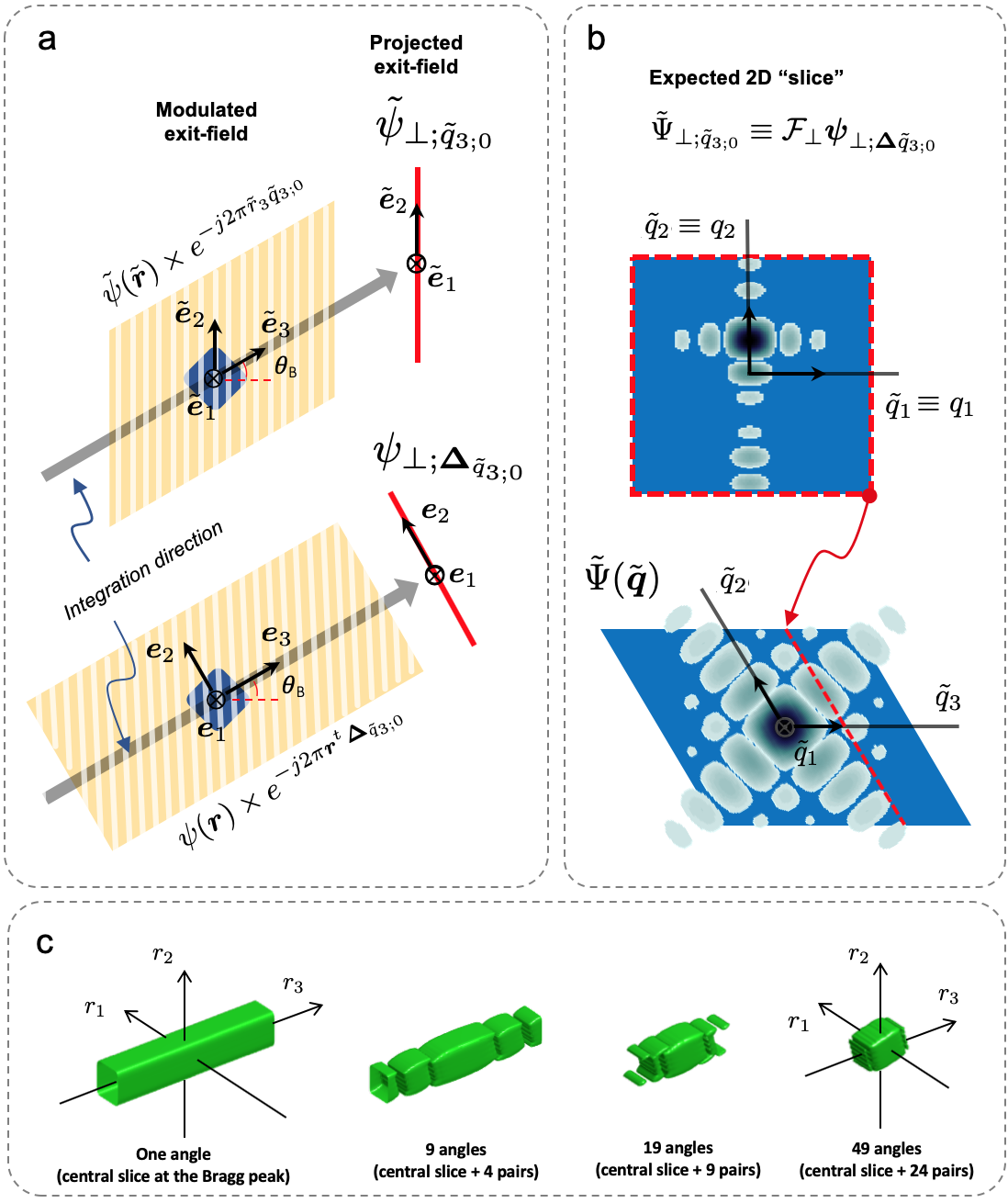}
\caption{Illustration of the projection/back-projection strategy. 
(a) The integrated exit-field in the non-orthogonal or in the 
orthogonal real-space representation allows to derive (b) a 
``slice'' $\tilde{\Psi}_{\bot;  \tilde{q}_{3;0}}$ extracted from  
$\tilde{\Psi}(\tilde{\qb})$ at $\tilde{q}_{3} =\tilde{q}_{3;0}$. (c) Conversely, the summation of a series of 
back-projections allows to retrieve the object from its 
evenly-sampled ``slices''; here, the retrieved  sample is 
converging toward the cubic, real-valued particle shown in 
Fig.~\ref{Fig:ReconstructionSquare3DFT}.}
\label{Fig:ProjBackProj}
\end{figure}

To illustrate some of the concepts and relationships of the projection/back-projection approach, we refer to Fig.~\ref{Fig:ProjBackProj}.
Panel (a) shows a cubic crystal expressed in two frames corresponding to the real-space conjugate measurement frame $\tilde{\eb}$ as well as the orthogonal sample frame $\eb$. 
As we have described, in both cases, different phase modulation terms should be applied. These phase modulations are represented in the Figure by the white/yellow stripes, and we find that they manifest themselves differently with respect to the two frames, but they both encode equivalent information. 
Thus, applying the integration along $\tilde{\eb}_3$ or
  $\eb_3$ respectively produces 2D exit wavefields that are the same.
Applying a 2D Fourier transform to the resulting projected exit field will produce the far field diffraction pattern shown in Fig.~\ref{Fig:ProjBackProj}(b).
This diffraction pattern corresponds to a slice of the overall 3D Fourier far field diffraction that is offset from the center by a desired amount along $\tilde{q}_3$.
The inverse process of constructing a description of the object $\psi$ from a series of backprojections corresponding to a set or $\tilde{\Psi}$ is shown in Fig.~\ref{Fig:ProjBackProj}(c). 
From a single slice, only the outline of the cube can be discerned. 
However, as real space backprojections from an increasing number of Fourier slices are integrated, the form of the cube takes shape. 
In the next section, we derive convenient means of numerically implementing these relations.

\subsection{Numerical evaluation with digital Fourier transforms}
\label{numerical_implementation_BP}
The numerical implementation of mapping between non-orthogonal representation spaces presented in 
Sec.~\ref{nonOrtgogonalProjBProj} is considered first. 
We note that the discussion presented here relies on some of the notation introduced 
in Sec.~\ref{numerical_implementation}.  
Let us introduce 
$\tilde{\psib} \in \eC^{N_2 \times N_1 \times N_3}$ as the 3D array built
from the regular spatial sampling of $\tilde{\psi}$, and $\tilde{\Psib}_{\bot; \tilde{q}_{3;0}} \in \eC^{N_2
    \times N_1}$ as the 2D array built from the regular sampling of the
  detector plane ($\tilde{\qb}_\bot$) at $\tilde{q}_{3}=\tilde{q}_{3;0}$.
We deduce from \eqref{sep_FT_2} the discretized form of the relationship:
\begin{equation}
  \label{numericalQuantities_NonOrtho1}
  \begin{array}{rcl}
    \tilde{\Psib}_{\bot; \tilde{q}_{3;0}} & = & \textbf{DFT}_\bot \big(
                                            \textbf{SUM}_3 \big(
                                            \tilde{\psib}
                                            \odot \tilde{\mub}_{\tilde{q}_{3;0}}
                                                  \big)\big) \times \delta_{\tilde{r}}
    \end{array}
\end{equation}
where $\delta_{\tilde{r}} \equiv \delta_{\tilde{r}_1}\delta_{\tilde{r}_2}\delta_{\tilde{r}_3}$ is the voxel volume in the non-orthogonal
frame $(\tilde{\eb}_1, \tilde{\eb}_2, \tilde{\eb}_3)$ and $\textbf{SUM}_3$ is
a summation (\textit{i.e.}, a projection)
operator acting along the third dimension of the 3D array. 
The array describing the phase modulation needed to produce the desired ``slice'' $\tilde{q}_{3}=\tilde{q}_{3;0}$ is given by:
\begin{equation}
  \label{numericalQuantities_NonOrtho2}
  \begin{array}{rcl}
    \tilde{\mub}_{\tilde{q}_{3;0}} & \equiv & \{  e^{-\iota 2\pi n_3 \delta_{\tilde{r}_3} \tilde{q}_{3;0}}, \, \forall (n_1,n_2) \} \in \eC^{N_2 \times N_1 \times N_3}. 
    \end{array}
\end{equation}
Conversely, the numerical implementation of the back-projection step \eqref{backprop} is given by: 
\begin{equation}
  \label{numericalQuantities_NonOrtho4}
    \tilde{\psib}_{\tilde{q}_{3;0}} 
     =  
                \tilde{\mub}_{\tilde{q}_{3;0}}^* 
                \odot
                \textbf{REP}_3 \big(
                \textbf{DFT}_\bot^{-1} \big(
                \tilde{\Psib}_{\bot;
                \tilde{q}_{3;0}} \big) \big) \times
              \frac{1}{\delta_{\tilde{r}_1}\delta_{\tilde{r}_2}}.
\end{equation}
In this expression, $\textbf{REP}_3$ is a replication (\textit{i.e.}, 
a back-projection) operator that creates an $ N_2 \times N_1 \times N_3$ 
array from a $N_2 \times N_1$ array, ensuring that the array is self-similar along $N_3$.  
In the back-projection step, the phase modulation term is the complex conjugate ($^*$) of $\tilde{\mub}_{\tilde{q}_{3;0}}$.

The sampling along the RC implies that a set of ``slices''
$\tilde{\Psib} \equiv \{\tilde{\Psib}_{\bot;  \tilde{q}_{3}} \,|\, \tilde{q}_3
=\tilde{q}_{3;0}, \cdots, \tilde{q}_{3;N_3-1}\}$ should be 
computed to provide the expected measurements that will be eventually
constrained by the data during phase retrieval. 
However, if we consider the case when the RC is regularly sampled, then the regularity of the mesh along $\tilde{\kb}_3$ 
allows one to sidestep the slice-by-slice approach and simply obtain all the slices with a single, much faster, 3D DFT:  $\tilde{\Psib} = \textbf{DFT}(\tilde{\psib})$.
As a result, the projection/back-projection strategy described here will be appealing in situations, for example when $\tilde{q}_3$ is unevenly-sampled, 
as was demonstrated in \cite{calvo-almazan19}.

We now consider the numerical implementation of the projection/back-projection approach in mapping between orthogonal sample space to a sheared measurement space, and back. 
Starting from the
orthogonal representation $\psi$, we deduce 
from \eqref{ProjGen} that a given far-field diffraction slice can be obtained numerically by:
\begin{equation}
  \label{numericalQuantities_Ortho1}
  \begin{array}{rcl}
    \tilde{\Psib}_{\bot; \tilde{q}_{3;0}} & = & \textbf{DFT}_\bot \big(
                                            \textbf{SUM}_3 \big(
                                            \psib
                                            \odot \mub_{\tilde{q}_{3;0}} 
                                                  \big)\big) \times
                                            \delta_{r} \cos \theta_{\text{\tiny B}}.
    \end{array}
\end{equation}
In this expression, $\psib\in \eC^{N_2\times N_1\times  N_3}$ is the 3D 
array built from the regular sampling of the orthogonal
real-space sample representation, and 
\begin{equation}
  \label{numericalQuantities_Ortho2}
  \begin{array}{rcl}
    \mub_{\tilde{q}_{3;0}} & \equiv & \{  e^{-\iota 2\pi \nb^T \Lambdab_r \Deltab_{\tilde{q}_{3;0}}}\} \in \eC^{N_2 \times N_1 \times N_3}
    \end{array}
\end{equation}
is the phase modulation array (the notations $\nb$ and $\Lambdab_r$ were introduced in 
Sec.~\ref{numerical_implementation}).  Conversely, the numerical implementation 
of the back-projection step \eqref{backproportho} is given by:
\begin{equation}
  \label{numericalQuantities_Ortho3}
    \psib_{\tilde{q}_{3;0}} 
     =  
                \mub_{\tilde{q}_{3;0}}^* 
                \odot
                \textbf{REP}_3 \big(
                \textbf{DFT}_\bot^{-1} \big(
                \tilde{\Psib}_{\bot;
                \tilde{q}_{3;0}} \big) \big) \times
          \frac{1}{\delta_{r_1}\delta_{r_2} \cos \theta_{\text{\tiny B}}}.
\end{equation}
Following Eq.~\eqref{backproportho}, the real-space back-projections 
corresponding  to a set of individual ``slices'' can be summed-up to provide  an estimate of the diffracting sample:
\begin{equation}
  \label{numericalQuantities_Ortho4}
    \psib 
     \approx  \sum_{n = 0}^{N_3-1} \psib_{\tilde{q}_{3;n}} \times
     \frac{1}{N_3 \delta_{r_3}}. 
\end{equation}

We note that Equations \eqref{numericalQuantities_Ortho1} and 
\eqref{numericalQuantities_Ortho3} can be used as flexible building blocks by which to design new
 phase retrieval reconstruction algorithms, as has been recently shown
\cite{hruszkewycz17,hill18,calvo-almazan19}. 
As an aid to interested readers, an example of \textsc{matlab} code for such an 
algorithm is provided  in Appendix~B in which  the 
projection/back-projection strategy is implemented for the BCDI-ER 
phase-retrieval algorithm. 

From a computational perspective,  we point out that the increased adaptability of the projection/back-projection approach does come at a price regarding computation speed. 
If we consider the situations in which slices are regularly sampled along the RC direction, then the repeated calculations of projection/back-projection over a regular mesh 
along $\tilde{q}_3$ is likely to be much slower than the 3D DFTs involved 
in \eqref{psi_num}. This is evident by noting that the run time of the ER phase retrieval code of the 3D DFT method (Algo.~\ref{ER_script} ) is ten times faster than that for projection/back-projection (Algo.~\ref{ER_BP_script}).

\section{Conclusion} 

Bragg CDI provides the ability to measure a volume of Fourier space containing rich structural information about a Bragg-diffracting sample, and to invert this measurement into a real-space image. 
The nature of the measurement is such that parallel sequential slices of Fourier space are obtained, which allows for the efficient and robust design of phase retrieval algorithms for this purpose. 
However,  using currently available phase retrieval tools, as-reconstructed Bragg CDI reconstructions exhibit geometric 
distortions induced by the non-orthogonal nature of the measurement 
in Fourier space. Although well understood, these distortions 
are a serious hurdle for the interpretation of the raw reconstruction.

In this two-part work, we provide a mathematically
comprehensive view of this problem, 
and we outline several strategies aimed to address the issue in different ways. 
In Part I,  we derive the means by which to transform the final 
raw BCDI reconstruction so that it can be displayed within 
an arbitrary orthogonal real-space frame. 
This approach currently in use in the BCDI community, and we provide in Part I the mathematical underpinnings of the method that will facilitate its adoption at a broader range of synchrotron beamlines.  
Part II addresses the problem from a different viewpoint. 
We aimed to develop Fourier transformations that are designed to reconcile the non-orthogonal frame of the measurement with an orthogonal representation of the sample. 
In this spirit, we derive two different transformations that achieve this for the case of regularly sampled and irregularly sampled experimental data sets. 
Importantly, these transformations can be embedded within the iterative loops of popular BCDI phase retrieval routines, so that a natural orthogonal sample frame is built into the phase retrieval framework \emph{a-priori}. 
This is tremendously advantageous when incorporating physical constraints into the design of a phase retrieval algorithm that stem from the experimental geometry. 
Thus, the work presented in Parts I and II together provide a unified set of concepts for the BCDI community to address issues related to achieving natural 3D reconstruction representation, and more broadly, to embed these concepts more deeply into phase retrieval in order to enable the design of new types of BCDI experiments that cannot be realized with current methods.

\appendix
\section{Extension to arbitrary, non-specular Bragg reflections}
\label{Sec:nonspec}
In most BCDI diffraction configurations, the ``orthogonal'' Fourier-space frame $(\kb_1,\kb_2, \kb_3)$ can be
chosen  such that $\kb_\bot \equiv (\kb_1, \kb_2) $ is identical to
the pair $\tilde{\kb}_\bot \equiv (\tilde{\kb}_1 ,\tilde{\kb}_2)$ that corresponds to 
the detection plane. As a consequence, a general form for the transformation 
matrix \eqref{basis_RQ_ortho} linking $\Psi$  and $\tilde{\Psi}$ is  
\begin{equation}
  \label{basis_RQ_ortho_App1}
       \Bb_{q} = 
       \left(
        \begin{array}{ccc}
          1   & 0& |\\
          0   & 1&\bb\\
          0 & 0& |
        \end{array}
       \right)
\end{equation}
In this expression, $\bb \in \eR^3$ is the decomposition of $\tilde{\kb}_3$ in 
the ``orthogonal'' basis $(\kb_1, \kb_2, \kb_3)$. The specific
expression of $\bb$  varies significantly in practice 
because it depends on the specific design of the diffractometer used in the measurement and on the nature of the scan.
One such example is given in Sec.~5 in the Part I companion paper. 
The transfomrations presented  
in Sec.~\ref{mapping}  and \ref{ProjBAckProj}  of this paper are
nevertheless easy to adopt to accommodate the generic form in 
\eqref{basis_RQ_ortho_App1}. 

First, let us consider the 
main results in  Sec.~\ref{ProjBAckProj}.  The relations \eqref{ProjGen}, 
\eqref{backproportho}, \eqref{numericalQuantities_Ortho1} and 
\eqref{numericalQuantities_Ortho3} hold under the adaptation considered here, 
provided that  the Fourier-space shift introduced in \eqref{shift}
reads
\begin{equation}
  \label{shiftApp}
  \Deltab_{\tilde{q}_{3;0}} \, = \, \tilde{q}_{3;0}   \, \bb.
\end{equation}
We specify that the form of $\bb$ is the following, conforming to the convention presented in Sec.~\ref{mapping}:
\begin{equation}
  \label{basis_RQ_ortho_App1a}
	\bb = [-b_1~~-b_2~~b_3]^T.
\end{equation}
Then, the expression for the ``forward'' step $\psi \rightarrow \tilde{\Psi}$ in the more general diffraction geometry in the continuous domain (analogous to relation \eqref{final_forward}) will be as follows:
\begin{equation}
  \label{OmegaApp1}
        \begin{array}{rcl}
          \tilde{\Psi}(\tilde{\qb}) &:=& b_3\times [\mathcal{F}_\bot
                                     \xi](\tilde{\qb}_\bot, b_3\tilde{q}_3) \\[1em]
          \text{with} \quad \xi(\rb_\bot, q_3) &:=& [\mathcal{F}_3
                                                    \psi] (\rb_\bot, q_3)
                                  \times e^{\iota 2\pi (\tilde{b}_1 r_1 + \tilde{b}_2 r_2) q_3 }
        \end{array}
\end{equation}
where $\tilde{b}_1\equiv \frac{b_1}{b_3}$ and $\tilde{b}_2\equiv
\frac{b_2}{b_3}$.
We can also update the counterpart expression in the discrete domain (analogous to  \eqref{psi_num}):
\begin{equation}
  \label{OmegaNumApp1}
        \begin{array}{rcl}
          \tilde{\Psib} &=& \, \textbf{DFT}_\bot\big(\textbf{DFT}_x (\psib) \, \odot \, \mub \big) \times b_3\delta_{r}  \\[1em]
          \text{with} \quad \mub & := & \{  e^{\iota 2\pi  (n_1 \delta_{r_1} \tilde{b}_1 + n_2
                          \delta_{r_2} \tilde{b}_2) \times
                      m_3 \delta_{\tilde{q}_3} b_3 } \}
        \end{array}
\end{equation}
and where $m_3\in \eZ$ and $(n_1,n_2)\in \eZ^2$. 
The details of deriving the vector $\bb$ require a detailed knowledge of the sample goniometer motions and of the angular degrees of freedom of the detector. 
An example of the derivation of $\bb$ that could be used here is given in Part I, emulating the experimental setup at the  34-ID-C coherent Bragg diffraction end station at the Advanced Photon Source.

\section{Matlab codes}
This appendix provides the \textsc{matlab} code of the modified ER phase retrieval algorithms
that reconstruct an image of the sample in the orthogonal real-space frame $(\eb_1, 
\eb_2, \eb_3)$.  The code in Algo.~\ref{ER_script} is 
derived from Sec.~\ref{numerical_implementation} and allows the 
3D intensity data stack to be transformed at once, applicable to the case when the RC is evenly sampled. 
The projection/back-projection version of ER given in Algo.~\ref{ER_BP_script}.
This approach allows an arbitrary set of angles to be used for phase retrieval, and employs the framework introduced in Sec.~\ref{numerical_implementation_BP}.

The aim of the codes presented in this Appendix is to illustrate that the modified ER 
code is not very different from the one currently in use for BCDI phase retrieval. The main difference is that
the commands \texttt{fftn/ifftn} used to link the sample to the 
3D field in the Fraunhofer regime are replaced by a different pair of functions. 
For example, these functions are \texttt{Real2NOF()} and
\texttt{NOF2Real()} in Algo.~\ref{ER_script}.  
In the example codes, the number of mesh points $N_1$ and $N_3$ are the number of physical pixels extracted from the
detector in $q_1$ and the number of angles in the RC, respectively. For the 
sake of simplicity, we did not consider the extent of the  Fourier 
domain in $q_2$ discussed in Sec.~\ref{numerical_implementation}.
As a result, $N_2$ is identical to the number of physical pixels
extracted from the detector in $q_2$. The origin of real and 
Fourier space are set to correspond to the central pixel of the respective numerical 
windows, which indicates that \texttt{fftshift}/\texttt{ifftshift} have to be used in combination with the Fourier transformations.
   
Finally, we note that other standard phase-retrieval algorithms such as HIO and 
Shrink-wrap can be generated easily by adapting the ER codes provided. Source codes
are available on request or can be downloaded directly from the 
repository given in \cite{CodesPartII19}.

\ack{
	The development and simulations of the shear-aware Fourier transformation formalism for regularly sampled rocking curves based on the discrete Fourier transformation was supported by the European Research Council (European Union's Horizon H2020 research and innovation program grant agreement No 724881).
The development of the back-projection-based Fourier transformation formalism for irregularly sampled rocking curves was supported by the U.S. Department of Energy, Office of Science, Basic Energy Sciences, Materials Science and Engineering Division.
 }


\begin{table}
\label{ER_script}
     \caption{--- Matlab script for the Error-reduction strategy retrieving the  orthogonal
       representation of the exit-field $\psi$ from the intensity of the non
       orthogonal representation of the 3D diffracted field
       $\tilde{\Psi}$.}
 \hrule \hrule\vspace{.5em}
{\color{gray}
\textbf{Required:}
 \begin{itemize}
 \item[] \texttt{dp} \text{: stack of intensity patterns [array: $N_2\times N_1 \times N_3$]}\\[-2em] 
 \item[]\texttt{psi} \text{: starting estimate [array: $N_2\times N_1 \times N_3$]}\\[-2em] 
 \item[]\texttt{supp} \text{: support of the 3D exit-field in the orthogonal frame [array: $N_2\times N_1 \times N_3$]}\\[-2em] 
 \item[]\texttt{dq1}, \texttt{dq2}, \texttt{dq3} \text{: Fourier-space sampling rates along $\kb_1$, $\kb_2$ and $\kb_3$ [scalars]}\\[-2em] 
 \item[]\texttt{thetaB} \text{: Bragg angle  [scalar]}\\[-2em]  
\end{itemize}
\textbf{Result:} 
\begin{itemize}
 \item[] \texttt{psi} \text{: retrieved sample in the ORTHOGONAL
   representation  [array: $N_2\times N_1 \times N_3$].}
\end{itemize}
}
\hrule  \vspace{.5em}
\begin{tabular}{l}
\%\%  \textbf{ER script}\\
\texttt{[N2,N1,N3]                  = size(dp);}\\
\texttt{dr1                             = 1/(N1*dq1);}
\texttt{dr2                             = 1/(N2*dq2);}
\texttt{dr3                             = 1/(N3*dq3);}\\
\texttt{r1                              = 0:dr1:(N1*dr1);}
\texttt{r2                              = 0:dr2:(N2*dr2);}
\texttt{r3                              = 0:dr3:(N3*dr3);}\\
\texttt{q1                              = 0:dq1:(N1*dq1);}
\texttt{q2                              = 0:dq2:(N2*dq2);}
\texttt{q3                              = 0:dq3:(N3*dq3);}\\
\texttt{[R1,R2,R3]                  = meshgrid(r1,r2,r3);}
\texttt{[Q1,Q2,Q3]                = meshgrid(q1,q2,q3);}\\[.25em]
\%\%  \textbf{Pre-computation of the 3D modulation array}\\
\texttt{phase\_ramp          = exp(i*2*pi*R2.*Q3*tan(thetaB));} \\
\texttt{param       = [dr1, dr2, dr3, thetaB];}\\[.25em]
\%\%  \textbf{The ER main loop}\\
\texttt{iter\_num = 10;} \hspace{1em} \% \text{Total  number of ER update}\\          
\texttt{alpha  = 1;} \hspace{1em} \% \text{Updating stepsize for  ER}\\[.25em]
\textbf{for} $\texttt{iter}   =1:\texttt{iter\_num}$\\
  \hspace{1em} \% Forward calculation: expected 3D far-field\\
  \hspace{1em} \texttt{PSI\_non\_ortho       =        Real2NOF (psi, phase\_ramp, param);}\\
  \hspace{1em} \% Modulus constraint in Fourier space:\\
  \hspace{1em} \texttt{PSI\_non\_ortho       =      sqrt(dp).*exp(1i*angle(PSI\_non\_ortho));}\\
  \hspace{1em} \% backward calculation: corrected exit-field\\
  \hspace{1em} \texttt{psi\_new       =       NOF2Real (PSI\_non\_ortho,  phase\_ramp,       param);}\\
  \hspace{1em} \% Real-space support constraint\\
  \hspace{1em} \texttt{psi    = psi - alpha*supp.*(psi -  psi\_new);}\\
\textbf{end}
\end{tabular}
\hrule
\begin{tabular}{l}
  \%\%  \textbf{Forward and backward transformation functions...}\\[.25em]
  \texttt{{\bf function}  PSI\_non\_ortho       =  Real2NOF(psi, phase\_ramp, param)}\\
  \texttt{dr1                = param(1);} 
  \texttt{dr2                = param(2);}  
  \texttt{dr3                = param(3);} 
  \texttt{thetaB               = param(4);}\\
  \texttt{scaling              = cos(thetaB)*dr1*dr2*dr3;}\\
  \texttt{XI\_rx         = fftshift(fft(psi,[],3),3).*phase\_ramp*scaling} \\
  \texttt{PSI\_non\_ortho       =  fftshift(fft(fftshift(fft(XI\_ortho\_rx,[],1),1),[],2),2)};\\ 
  \texttt{end}\\[.25em]
  \texttt{{\bf function} psi       =  NOF2Real (PSI\_non\_ortho, phase\_ramp, param)}\\
      \texttt{dr1                = param(1);}
      \texttt{dr2                = param(2);} 
      \texttt{dr3                = param(3);} 
      \texttt{thetaB             = param(4);}\\
      \texttt{scaling            = cos(thetaB)*dr1*dr2*dr3;}\\  
      \texttt{zeta\_r3       =      fftshift(ifft(ifft(PSI\_non\_ortho,[],1),[],2),3).*conj(phase\_ramp);}\\ 
      \texttt{psi           = ifft(ifftshift(zeta\_rx,3),[],3)/scaling;}\\ 
      \texttt{end}
\end{tabular}
\hrule\hrule
\end{table}

\begin{table}
\label{ER_BP_script}
\caption{--- Idem Algo.~\ref{ER_script} but with projection and
  backprojection operators. We consider in this script that the RC is
  evenly sampled, with a sampling rate about $q_3$ that meets the
  requirement $\delta_{q_3} = \delta_{\tilde{q}_3} \cos
  \theta_{\text{\tiny B}}$, see \eqref{sampling_QS_x}.}
 \hrule \hrule\vspace{.5em}
{\color{gray}
\textbf{Required:}
 \begin{itemize}
 \item[] \texttt{dp} \text{: stack of intensity patterns [array: $N_2\times N_1 \times N_3$]}\\[-2em]  
 \item[]\texttt{psi} \text{: starting estimate [array: $N_2\times N_1 \times N_3$]}\\[-2em] 
 \item[]\texttt{supp} \text{: support of the 3D exit-field in the orthogonal frame [array: $N_2\times N_1 \times N_3$]}\\[-2em] 
 \item[]\texttt{dq1}, \texttt{dq2}, \texttt{dq3} \text{: sampling rates along $\kb_1$, $\kb_2$ and $\kb_3$ [scalars]}\\[-2em] 
\item[] \texttt{thetaB} \text{: Bragg angle  [scalar]}\\[-2em]  
\end{itemize}
\textbf{Result:} 
\begin{itemize}
 \item[] \texttt{psi} \text{: retrieved sample in the ORTHOGONAL
   representation  [array: $N_2\times N_1 \times N_3$].}
\end{itemize}
}
\hrule \hrule \vspace{.5em}
\begin{tabular}{l}
\%\%  \textbf{ER script (project/back-projection version)}\\
\texttt{[N2,N1,N3]                  = size(dp);}\\
\texttt{dr1                             = 1/(N1*dq1);}
\texttt{dr2                             = 1/(N2*dq2);}
\texttt{dr3                             = 1/(N3*dq3);}\\
\texttt{r1                              = 0:dr1:(N1*dr1);}
\texttt{r2                              = 0:dr2:(N2*dr2);}
\texttt{r3                              = 0:dr3:(N3*dr3);}\\
\texttt{q1                              = 0:dq1:(N1*dq1);}
\texttt{q2                              = 0:dq2:(N2*dq2);}
\texttt{q3                              = 0:dq3:(N3*dq3);}\\
\texttt{[R1,R2,R3]                  = meshgrid(r1,r2,r3);}\\
\texttt{[Q1,Q2,Q3]                = meshgrid(q1,q2,q3);}\\
\texttt{scaling                       = cos(thetaB)*dr1*dr2;}\\[.5em]
\%\%  \textbf{The ER main loop}\\
\texttt{iter\_num = 10;} \hspace{1em} \% \text{Total  number of ER update}\\          
\texttt{alpha  = 1;} \hspace{1em} \% \text{Updating stepsize for  ER}\\[.25em]
   \textbf{for} \texttt{iter   = 1 : iter\_num} \\[.25em]
       \hspace{1em}   \texttt{psi\_ortho\_new           = 0;}\\[.25em]
       \hspace{1em}   \% \text{Loop over the RC angles...}\\[.25em]        
           \hspace{1em}  \textbf{for} \texttt{n\_q3 = 1:N3} \\[.25em]             
             \hspace{2em}  \texttt{sqrtI               =   squeeze(dp\_sqrt(:,:,n\_q3));}\\[.25em]     
             \hspace{2em}   \% \text{Computation of the current slice in the 3D expected far-field...}\\            
             \hspace{2em}   \texttt{phase\_ramp          = exp(1i*2*pi*q3(n\_q3)*(R3 - tan(theta\_B)*R2));}\\           
             \hspace{2em}   \texttt{psi\_mod\_integ       = squeeze(sum(psi\_ortho.*conj(phase\_ramp),3))*dr3;}\\
             \hspace{2em}   \texttt{PSI\_non\_ortho       = fftshift(fft2(fftshift(psi\_mod\_integ)))*scaling;}\\[.25em]      
             \hspace{2em}   \% \text{Contribution to the 3D corrected  far-field}\\
             \hspace{2em}   \texttt{PSI\_non\_ortho       = sqrtI.*exp(1i*angle(PSI\_non\_ortho));}\\
             \hspace{2em}   \texttt{psi\_mod\_integ       = ifftshift(ifft2(ifftshift(PSI\_non\_ortho)))/scaling;}\\
             \hspace{2em}   \texttt{psi\_ortho\_new       =              psi\_ortho\_new + ... }\\
             \hspace{6em}   \texttt{repmat(psi\_mod\_integ, [1,1,length(q3)]).*phase\_ramp/(N3*dr3);}\\[.25em]                
            \hspace{1em} \textbf{end}\\[.25em]
        \hspace{1em}  \% \text{Apply the support constraint...}\\
        \hspace{1em} \texttt{psi\_ortho           = psi\_ortho - alpha*supp.*(psi\_ortho - psi\_ortho\_new);}\\[.25em]
    \textbf{end}
\end{tabular}
\end{table}

\end{document}